\documentclass[a4paper]{article}
 
\usepackage{graphics}
\usepackage{adjustbox,lipsum}
\usepackage[english]{babel}
\usepackage[utf8]{inputenc}
\usepackage{amsmath}
\usepackage{graphicx}
\usepackage{hyperref}
\usepackage{url}
\usepackage{subcaption}
\usepackage[colorinlistoftodos]{todonotes}
\usepackage{ upgreek }
\usepackage{amsfonts}
\usepackage{caption}
\usepackage{float}
\usepackage{subcaption}
\usepackage{multirow}

\title{Bayesian Quantile Regression Using Random B-spline Series Prior}

\author{Priyam Das\footnote{Deaprtment of Statistics, North Carolina State University, NC, USA} $\/$ and Subhashis Ghoshal \footnote{Deaprtment of Statistics, North Carolina State University, NC, USA}}

\date{\today}

\begin{document}
\maketitle
\begin{abstract}
We consider a Bayesian method for simultaneous quantile regression on a real variable. By monotone transformation, we can make both the response variable and the predictor variable take values in the unit interval. A representation of quantile function is given by a convex combination of two monotone increasing functions $\xi_1$ and $\xi_2$ not depending on the prediction variables. In a Bayesian approach, a prior is put on quantile functions by putting prior distributions on $\xi_1$ and $\xi_2$. The monotonicity constraint on the curves $\xi_1$ and $\xi_2$ are obtained through a spline basis expansion with coefficients increasing and lying in the unit interval. We put a Dirichlet prior distribution on the spacings of the coefficient vector. A finite random series based on splines obeys the shape restrictions. We compare our approach with a Bayesian method using Gaussian process prior through an extensive simulation study and some other Bayesian approaches proposed in the literature. An application to a data on hurricane activities in the Atlantic region is given. We also apply our method on region-wise population data of USA for the period 1985--2010.
\end{abstract}

\section{Introduction}

 In a regression model, if the distribution of the response variable is highly skewed, traditional mean regression model may fail to describe interesting aspect of the relationship between the prediction and response variables. For instance, if we deal with the income distribution data of a certain state or country, the traditional mean regression will get affected by the outliers, i.e., the income of top 1--2 \% people of that region, and hence is of limited use for prediction of income of general people. This situation arises often in business, economics, environmental and many other fields. As an alternative to traditional linear regression, quantile regression is one of the most popular and useful regression technique. 
 
 Extensive research have been done on quantile regression. Most of them were approached \quad from \quad the \quad frequentist \quad perspective (\cite{Koenkar2005}). 
 \cite{Koenkar1978} estimated the $\tau$th regression quantile for the dataset $\{(X_i,Y_i)\}_{i=1}^n$, by minimizing the loss function $\sum_{i=1}^n (Y_i-\beta_0-\beta_1 X_i)[\tau-I(Y_i-\beta_0-\beta_1 X_i<0)]$ with respect to $\beta_0$ and $\beta_1$. Due to computational convenience and other theoretical properties, this method remained popular for long time. Later on, many other methods of quantile regression were proposed both in frequentist and Bayesian framework. \cite{Yu2001} introduced Bayesian quantile regression for independent data. A few generalization and extension of quantile regression were proposed in \cite{Kottas2001}, \quad \cite{Gelfand2003},  \quad \cite{Kottas2009}. These \; papers mainly focused on quantile regression for a single quantile level for censored independent data. \cite{Geraci2007} proposed the single level quantile regression model with subject specific random intercept term which accounts for within group correlation. The drawback of separate quantile regression for different quantiles is that the natural ordering among different quantiles is not ensured. In other words, the lower quantile estimation curves might cross the upper quantile curves which violates the ordering of quantiles of a fixed distribution.  
 
 There exist a few methods addressing the crossing issue in estimating multiple quantile regression. \cite{He1997} proposed a method of estimating non-crossing quantile curves assuming a heteroskedastic model for response variable. Under this assumption, predictors might affect the response variable via a location and scale change. \cite{Neocleous2008} proposed a method of estimating the quantile curve by linear interpolation on the estimated quantile curves on a grid of quantiles. \cite{Takeuchi2004} and \cite{Takeuchi2006} used support vector machines (SVM, \cite{Vapnik1995}) for non-crossing quantile regression. But, as mentioned in \cite{Shim2010}, a disadvantage of using SVM for non-crossing quantile regression is that when multiple quantiles are needed, every adjacent pair of conditional quantile functions should be computed. \cite{Shim2009} proposed non-crossing quantile regression using doubly penalized kernal machine (DPKM). For this method also, we need to compute quantile curves separately for each quantile. The method proposed in \cite{Wu2009} sequentially updates the quantile curves under the contraint that the upper quantile curves stay above the lower ones. A drawback of this method is that the predicted quantile curves are dependent on the grid of quantiles for which we want to calculate the quantile curves. 
 
Quantile regression for fixed number of levels of quantiles with monotonicity constraint was proposed in \cite{Dunson2005} and \cite{Liu2011}. The method proposed in \cite{Bondell2010} is sensitive to the number of prediction quantile levels. For spatial quantile regression \cite{Reich2011} used Bernstein polynomial to maintain the monotonicity constraint of the quantile function. \cite{Reich2012} proposed non-crossing multiple quantile regression using piece-wise Gaussian basis functions to interpolate the quantile function at different quantile levels. \cite{Reich2013} suggested the use of different symmetric and asymmetric quantile functions for modeling based on the data.
Models based on a fixed number of quantiles do not give the estimates of all quantiles and may also be sensitive to the number and the location of the grid points of the quantile levels, which is not desirable. 
 
 Instead of fitting quantile curves for a fixed set of quantiles, a more informative picture emerges by estimating the entire quantile curve. Suppose $Q(\tau|x)=\text{inf}\{q\, : \, P(Y\leq q| X=x)\geq \tau\}$ denote the $\tau$-th conditional quantile $(0\leq\tau\leq 1)$ of a response $Y$ at $X=x$, $X$ being the predictor. A linear simultaneous quantile regression model for $Q(\tau|x)$ at a given $\tau$ is given by $$Q(\tau|x)=\beta_0(\tau)+x\beta(\tau)$$ where $\beta_0(\tau)$ is the intercept and $\beta(\tau)$ is the slope smoothly varying as function of $\tau$. Thus estimating the quantile function involves non-parametric estimation of the function $\beta_0(\tau)$ and $\beta(\tau)$. The main challenge in fitting this kind of model remains in complying with the monotonicity restriction of the predicted quantile lines $\beta_0(\tau)+x\beta_1(\tau)$ as a function of all values of the predictor $X$. 
 
 \cite{Tokdar2012} obtained a very useful characterization of the monotonicity constraint (see Equation (\ref{eq:theorem}) below). They used the characterization to propose a suitable prior on the quantile function in a Bayesian approach, and computed the posterior distribution of the quantile curves. More specifically, they used two (possibly) dependent Gaussian processes $\xi_1(\cdot)$ and $\xi_2(\cdot)$ to induce prior on $\beta_0(\tau)$ and $\beta(\tau)$ in such a way so that $Q(\tau|x)$ remains a monotonically increasing function of $\tau$. 
 
 Gaussian processes do not have any shape restriction. To induce monotonicity property on $\xi_1(\cdot)$ and $\xi_2(\cdot)$, \cite{Tokdar2012} took the running integral of the exponential transformation of two (possibly correlated) Gaussian processes. To evaluate the likelihood, it requires solving the equation $Q(\tau|x) = y$ (see Equation (\ref{tau_x_y}) below). Without having a convenient expression, the solution needs to be done completely numerically which involves computing integrals of transformed Gaussian process over a very fine grid. Substantial improvement in computing approach is possible using a finite random series prior (\cite{Shen2015}). Especially using a B-spline basis, monotonicity can be ensured by choosing a prior only on coefficients in increasing order. Piece-wise polynomial representation of splines allows obtaining $\tau_x(y)$ solving analytically. Thus we can reduce computation time substantially using a random series prior based on the B-spline basis.

\section{Model Assumptions}
\label{section_model}

We observe $n$ independent random samples $(X_1,Y_1),\ldots,(X_n,Y_n)$ of an  
explanatory variable $X$ and a response variable $Y$, both of which are assumed to be univariate. By monotonic transformations, we transform both of them to the unit interval. From Theorem 1 of \cite{Tokdar2012}, it follows that a linear specification $\text{$Q$}(\tau|x)=\beta_0(\tau)+x\beta_1(\tau)$, $\tau \in [0,1]$, is monotonically increasing in $\tau$ for every  $x\in [0,1]$ if and only if
\begin{equation}\label{eq:theorem}Q(\tau|x)=\mu + \gamma x +\sigma_1x\xi_1(\tau)+\sigma_2(1-x)\xi_2(\tau),
\end{equation}
where $\sigma_1$ and $\sigma_2$ are positive constants and $\xi_1,\xi_2 : [0,1] \mapsto [0,1]$ are monotonically increasing in $\tau  $$\in$$[0,1]$. Because $Y$ also has domain $[0,1]$, we have the boundary conditions  
$Q(0|x)=0$ and $Q(1|x)=1$. Now, putting $\tau=0$ in Equation (\ref{eq:theorem}), and using $\xi_1(0)=\xi_2(0)=0$, we obtain  
$$\mu + \gamma x = 0 \quad \text{for all} \quad x \in [0,1]$$
which implies $\mu=\gamma=0$ and hence (\ref{eq:theorem}) reduces to 
\begin{equation}\label{eq:theorem_2}Q(\tau|x)=\sigma_1x\xi_1(\tau)+\sigma_2(1-x)\xi_2(\tau).
\end{equation}
Now putting $\tau=1$ and using $\xi_1(1)=\xi_2(1)=1$, we obtain 
$$1=\sigma_1x+\sigma_2(1-x) \; \text{for all} \; x \in [0,1],$$
which implies $\sigma_1=\sigma_2=1$. Therefore in the present context, the quantile regression function has representation 
\begin{align}
Q(\tau|x)= x\xi_1(\tau)+(1-x)\xi_2(\tau) \; \text{for} \; \tau \in [0,1], \; x,y \in [0,1].
\label{our_eq}
\end{align}
Equation (\ref{our_eq}) can be re-framed as
\begin{align}
Q(\tau|x)= \beta_0(\tau)+x\beta_1(\tau) \; \text{for} \; \tau \in [0,1], \; x,y \in [0,1],
\label{our_eq_slope_int}
\end{align}
where $\beta_0(\tau) = \xi_2(\tau)$ and $\beta(\tau) = \xi_1(\tau)-\xi_2(\tau)$ denotes the slope and the intercept of the quantile regression, which are smooth functions of $\tau$. Provided that $Q(\tau|x)$ is strictly increasing in $\tau$ for all $x$, the conditional density for $Y$ at $y$ given $X=x$ is given by 
 \begin{align}
 f(y|x) = \bigg(\frac{\partial}{\partial \tau}Q(\tau|x)|_{\tau=\tau_x(y)}\bigg)^{-1}=\bigg(\frac{\partial}{\partial \tau}\beta_0(\tau)+x\frac{\partial}{\partial \tau}\beta(\tau)|_{\tau=\tau_x(y)}\bigg)^{-1},
 \label{eq : density}
 \end{align}
 where $\tau_x(y)$ solves the equation 
 \begin{align}
x\xi_1(\tau)+(1-x)\xi_2(\tau) = y.
\label{tau_x_y}
 \end{align} 
 Then the joint conditional density of $Y_1,\ldots,Y_n$ given $X_1,\ldots,X_n$ is given by $\prod_{i=1}^n f(Y_i|X_i)$. 

\section{Regression with Spline}

Function estimation on a bounded interval through B-spline basis expansion is one of the most convenient approaches. To construct a prior on the quantile function $Q(\tau|x)$, or equivalently on $\xi_1$ and $\xi_2$, we need to ensure their monotonically increasing properties. If the coefficients in a B-spline basis expansion are in increasing order, then the corresponding quantile function will be an increasing function of $\tau$ (\cite{Boor2001}). In fact, if the target function is strictly increasing, the quality of the approximation of spline expansions is maintained when the coefficients are restricted by increasing order (\cite{Shen2015}). This motivates us to use B-splines with increasing coefficients as the basis of the quantile function. The degree of B-spline determines the degree of smoothness, in that quadratic splines are continuously differentiable, cubic splines are continuously twice differentiable and so on, and are able to optimally approximate functions of smoothness index only up to the degree of splines. Thus optimal estimation of smoother functions requires higher degree splines. On the other hand, computational complexity increases with the degree of the B-spline basis. Smoothness of order higher than two is typically not visually distinguishable from smoothness of order $2$, and therefore splines of degree up to 3 suffices in most applications. 
In case of quadratic B-spline basis, $Q(\tau|x)$ is a linear combination of piecewise second degree polynomials in terms of $\tau$, and hence restricted to an interval between two knots, $Q(\tau|x)$ is an increasing quadratic function. Thus we can solve $Q(\tau|X_i)=Y_i$ for each data points $(X_i, Y_i)$ analytically, unlike the case of Gaussian process prior, where one has to use numerically integrate and apply iterative Newton-Raphson method to solve them for each realization of the quantile process from its posterior distribution. Due to the increasing complexity and higher computation time, results for with B-splines of order higher than 3 have not been shown in this paper. For a cubic B-spline, $Q(\tau|x)$ is linear combination of piecewise third degree polynomials. Hence, for any given interval in the domain, $Q(\tau|x)$ is a monotonically increasing third degree polynomial. Hence in this case also, $\tau_x(y)$ can be solved analytically. However the analytic algorithm for solving cubic equation requires computing all three roots even though only one root falls in the admissible interval, which makes the cost of computing roots in a cubic equation substantial. Due to the monotonicity of $Q(\tau|x)$, the bisection method of finding roots is an attractive alternative. In our numerical experiments, we found that for cubic splines the bisection method is slightly faster than the analytical approach.

Let $0=t_0< t_1< \cdots < t_k=1$ be the equidistant knots on the interval $[0,1]$ where $t_i={i}/{k}$, $i=0,1,\ldots,k$. For B-spline of $m$th degree, the number of basis functions is $J=k+m$. Let $\{B_{j,m}(t)\}_{j=1}^{k+m}$ be the basis functions of $m$th degree B-splines on $[0,1]$ on the above mentioned equidistant knots. 
We consider the following basis expansion of the quantile functions through the relations
\begin{align}
& \xi_1(\tau)=\sum\limits_{j=1}^{k+m} \theta_j B_{j,m}(\tau) \; \text{ where } \; 0=\theta_1<\theta_2<\cdots<\theta_{k+m}=1,  \nonumber \\
& \xi_2(\tau)=\sum\limits_{j=1}^{k+m} \phi_j B_{j,m}(\tau) \; \text{ where } \; 0=\phi_1<\phi_2<\cdots<\phi_{k+m}=1.
\label{eq:constraint}
\end{align}
Taking $\theta_1=\phi_1=0$ ensures $\xi_1(0) = \xi_2(0)=0$ and $\theta_{k+m}=\phi_{k+m}=1$ ensures $\xi_1(1) = \xi_2(1)=1$. The increasing values of the coefficients takes care on monotonicity (\cite{Boor2001}). Thus, $\xi_1$ and $\xi_2$ are monotonically increasing functions of $\tau$ from $ [0,1]$ onto $ [0,1]$. Let, $\{\gamma_j\}_{j=1}^{k+m-1}$ and $\{\delta_j\}_{j=1}^{k+m-1}$ be defined by 
\begin{align}
\gamma_j=\theta_{j+1}-\theta_j, \; \delta_j=\phi_{j+1}-\phi_j, \quad j=1,\ldots,k+m-1. 
\label{eq : relation_3_0}
\end{align}
Hence 
\begin{align}
\theta_j=\sum_{i=1}^{j}\gamma_i, \; \phi_j=\sum_{i=1}^{j}\delta_i, \quad j=1, \ldots, k+m-1. 
\label{eq : relation_3}
\end{align}
and the restrictions in (\ref{eq:constraint}) equivalently be written as
\begin{align}
\gamma_j,\delta_j \geq 0,\; j=1,\ldots,k+m-1, \; \text{and} \; \sum_{j=1}^{k+m-1}\gamma_j= \sum_{j=1}^{k+m-1}\delta_j=1.
\label{eq : simplex}
\end{align}
Therefore the spacings of the B-spline coefficients take value in the unit simplex. A natural prior is thus given by the  uniform distribution on the unit simplex, or the Dirichlet prior with parameter $(1,\ldots,1)$. The Bayesian procedures based on quadratic splines (i.e. $m=2$) will be called the Quadratic Spline Simultaneous Quantile regression (QSSQR) and that based on cubic splines (i.e. $m=3$) will be called the Cubic Spline Simultaneous Quantile regression (CSSQR). 

\subsection{Model Fitting}
To compute the log-likelihood function, we proceed like \cite{Tokdar2012}. By (\ref{eq : density}), the log-likelihood is given by
\begin{align}
\sum\limits_{i=1}^n\log f(Y_i|X_i)&=-\sum\limits_{i=1}^n\log \frac{\partial}{\partial \tau}Q(\tau_{X_i}(Y_i)|X_i)\nonumber\\
&=-\sum\limits_{i=1}^n \log\Big\{X_i\frac{\partial}{\partial\tau}\xi_1(\tau_{X_i}(Y_i))+(1-X_i)\frac{\partial}{\partial\tau}\xi_2(\tau_{X_i}(Y_i))\Big\},
\end{align}
where $\tau_{X_i}(Y_i)$ is obtained from \eqref{tau_x_y}. 

Due to the monotonicity of $\xi_1$ and $\xi_2$, any convex combination of them will have one and only one solution on the interval $[0,1]$. To evaluate the log-likelihood we compute the derivatives of the basis splines (\cite{Boor2001}),
\begin{align*}
\frac{d}{dt}\xi_1(t) = \sum\limits_{j=2}^{k+m} \tilde\theta_j B_{j-1,m-1}(t), \quad 
\frac{d}{dt}\xi_2(t) = \sum\limits_{j=2}^{k+m} \tilde\phi_j B_{j-1,m-1}(t),
\end{align*}
where
\begin{align*}
\tilde\theta_j = (k+m)(\theta_j - \theta_{j-1}), \; \tilde\phi_j = (k+m)(\phi_j - \phi_{j-1}), \quad j = 2, \ldots,{k+m}. 
\end{align*}
The log-likelihood $\sum\limits_{i}\log f(Y_i|X_i)$  thus reduces to 
$$-\sum\limits_{i} \log\Big\{X_i\sum_{j=2}^{k+m}\theta^{\prime}_j B_{j-1,m-1}(\tau_{X_i}(Y_i)) 
  +(1-X_i)\sum_{j=2}^{k+m}\phi^{\prime}_j B_{j-1,m-1}(\tau_{X_i}(Y_i))\Big\}. 
$$

\subsection{MCMC and Transition Step}

In Equation (\ref{eq : simplex}), we note that $\{\gamma_j\}_{j=1}^{k+m-1}$ and $\{\delta_j\}_{j=1}^{k+m-1}$ are on the unit simplex. We shall use a Metropolis-Hastings algorithm to obtain Markov Chain Monte Carlo (MCMC) samples from the posterior distribution. In MCMC, to move on the simplex, we generate independent sequences $U_j$ and $W_j$, $j=1,\ldots,k+m-1$, from $U({1}/{r},r)$ for some $r>1$. 

Define $V_j=\gamma_j U_j$ and $T_j=\delta_j W_j$ for $j=1,\ldots,k+m-1$. Consider the proposal moves $\gamma_j \mapsto \gamma_j^*$ and $\delta_j \mapsto \delta_j^*$ given by 
\begin{align}
\gamma_j^* = \frac{V_j}{\sum_{j=1}^kV_j}, \; \delta_j^* = \frac{T_j}{\sum_{j=1}^kT_j}, \; i=1,\ldots,k+m-1. 
\end{align}
As shown in the appendix, the conditional distribution of $\{\gamma_j^*\}_{j=1}^{k+m-1}$ given $\{\gamma_j\}_{j=1}^{k+m-1}$ is 
\begin{align*}
g(\gamma^*|\gamma)=& \bigg(\frac{r}{r^2-1}\bigg)^{k+m-1}\bigg(\prod\limits_{j=1}^{k+m-1} \gamma_j\bigg)^{-1} (k+m-1)^{-1}\nonumber \\
&\quad\times\bigg[\Big\{\min_{0 \leq j \leq k+m-1}  ({r\gamma_j}/{\gamma_j^*})\Big\}^{k+m-1}-\Big\{\max_{0 \leq j \leq k+m-1} \frac{\gamma_j}{r\gamma_j^*}\Big\}^{k+m-1}\bigg].
\end{align*}
Similarly, we can find the conditional distribution of $\{\delta_j^*\}_{j=1}^{k+m-1}$. Now from the given set of values for $\{\gamma^*_j\}_{j=1}^{k+m-1}$ and $\{\delta^*_j\}_{j=1}^{k+m-1}$ the updated values $\{\theta^*_j\}_{j=1}^{k+m-1}$ and $\{\phi^*_j\}_{j=1}^{k+m-1}$ can be found using the following relation.
\begin{align}
\theta_j^*=\sum_{i=1}^{j}\gamma_i^*, \; \phi_j^*=\sum_{i=1}^{j}\delta_i^* \quad j=1, \ldots, k+m-1. 
\label{eq : relation_4}
\end{align}
While evaluating the likelihood, using the relations mentioned in Equation (\ref{eq : relation_3_0}) and (\ref{eq : relation_3}), we can find the likelihoods at the initial and destination points respectively in terms of $\{\gamma_j\}_{j=1}^{k+m-1}$ and $\{\delta_j\}_{j=1}^{k+m-1}$, and  $\{\gamma^*_j\}_{j=1}^{k+m-1}$ and $\{\delta^*_j\}_{j=1}^{k+m-1}$. The acceptance probability in Metropolis-Hastings algorithm is then given by 
\begin{align}
P_a&= \min \bigg\{\frac{L(\gamma^*,\delta^*)\pi(\gamma^*)\pi(\delta^*)f(\gamma|\gamma^*)f(\delta|\delta^*)}{L(\gamma, \delta)\pi(\gamma)\pi(\delta)f(\gamma^*|\gamma)f(\delta^*|\delta)},\; 1 \bigg\} \nonumber \\
&= \min \bigg\{\frac{L(\gamma^*,\delta^*)f(\gamma|\gamma^*)f(\delta|\delta^*)}{L(\gamma, \delta)f(\gamma^*|\gamma)f(\delta^*|\delta)},\; 1 \bigg\},
\end{align}
where $L(\cdot)$ denotes the likelihood and $\pi$ is the uniform Dirichlet prior density on the corresponding parameters.

\subsection{Choosing value of $k$ and Model Averaging}
\label{sec_model_avg}

In the previous sections, we discussed fitting spline basis functions for fixed number of partitions of equal length in $[0,1]$, the transformed domain of $X$. In other words, we developed the methodology for fixed number of basis functions. The number of basis functions $k+m$ controls the smoothness and hence the quality of the estimate. For smaller value of $k$, the bias is high but variability is less, so there is a bias-variance trade-off. Moreover MCMC chain runs faster with better mixing if the value of $k$ is smaller. It is desirable to determine the value of $k$ based on the data to apply the right amount of smoothness. In the Bayesian setting, it is natural to put a prior on the smoothing parameter $k$ and make inference based on the posterior distribution. For computational efficiency, the range of $k$ should be a finite collection of consecutive integers. Once we fix the range of all possible values of $k$, we can either consider empirical Bayes selection based on the marginal likelihood of $k$, or a Bayesian model averaging. A common approach to  posterior computation for a parameter space of varying dimension is through the reversible jump MCMC method, but its implementation can be challenging. An alternative approach is provided by the method described in \cite{Chib2001}. In this approach marginal probabilities corresponding to different values of $k$ are obtained from separate MCMC output for each value of $k$. Since all these chains are run independently of each other,  parallel computing can be implemented, which can greatly offset the cost of running separate chains for all feasible values of $k$.  Both the empirical Bayes approach by identifying the value of $k$ with highest posterior probability, and the hierarchical Bayes approach based on Bayesian model averaging, can be implemented using this method, and they have identical computing cost. 

Consider the $m$th degree B-spline with $k$ many partitions of $[0,1]$ of equal length. The equidistant knot sequence is given by $0=t_0<t_1<\cdots<t_k=1$ such that $t_i =i/k$ for $i =0,\ldots,k$. Let $\omega = (\{\gamma_j\}_{j=1}^{k+m-1},\{\delta_j\}_{j=1}^{k+m-1})$ be the parameter of interest. We denote the whole data by $z$, and with a slight abuse of notation, denote the joint density of $z$ also by $f$. The posterior density is then given by $$\pi(\omega|z)\propto\pi(\omega)f(z|\omega)$$ over $S$, a subset of $\mathbb{R}^{2(k+m-1)}$, specified by the Equation (\ref{eq : simplex}). For any $\omega^*$, the logarithm of the marginal likelihood is given by 
\begin{align}
\log{m}(z) = \log f(z|\omega^*) + \log \pi(\omega^*) - \log{\pi}(\omega^*|z).
\label{eq:marlik_exact}
\end{align}
It is recommended to take $\omega^*$ to be a point where it has high density under the posterior. For any given value of $\omega^*$, we can easily calculate the first two terms of the Equation (\ref{eq:marlik_exact}). Our goal is to estimate the posterior ordinate $\pi(\omega^*|y)$ given the posterior sample $\{\omega^{(1)}, \omega^{(2)}, \cdots, \omega^{(M)}\}$, where $M$ denotes the number of draws from the posterior sample after burn-in. Let us denote the proposal density by $q(\omega,\omega^{\prime}|z)$ for the transition from $\omega$ to $\omega^{\prime}$. Suppose the probability of move $\alpha(\omega,\omega^{\prime}|z)$ is given by $$\alpha(\omega,\omega^{\prime}|z) = \min\bigg\{1, \frac{f(z|\omega^{\prime})\pi(\omega^{\prime})q(\omega^{\prime},\omega|z)}{f(y|\omega)\pi(\omega)q(\omega,\omega^{\prime}|z)}\bigg\}.$$ \cite{Chib2001} proposed a simulation consistent estimate of the posterior ordinate given by 
\begin{align}
\hat{\pi}(\omega^*|z) = \frac{M^{-1}\sum_{g=1}^M\alpha(\omega^{(g)}, \omega^*|z)q(\omega^{(g)}, \omega^*|z)}{L^{-1}\sum_{j=1}^L\alpha(\omega^*, \tilde{\omega}^{(j)}|z)},
\label{eq:chib_expression}
\end{align}
where $\{\omega^{(g)}\}$ are the sampled draws from the posterior distribution and $\{\tilde{\omega}^{(j)}\}$ are the draws from $q(\omega^*,\tilde{\omega}|z)$, given the fixed value $\omega^*$. Then the estimated logarithm of the marginal is given by
\begin{align}
\log\hat{m}(z) = \log f(z|\omega^*) + \log \pi(\omega^*) - \log \hat{\pi}(\omega^*|z).
\label{eq:marlike}
\end{align}
As mentioned in \cite{Chib2001}, the computation time of the marginal likelihood of each of the probable model is small and is almost readily available after the MCMC chain finishes. After finding the marginal likelihood for each of those models, the value of $k$ corresponding to the model with the highest marginal likelihood can be treated as the best possible value of $k$. Suppose, we fix the domain of $k$ to be $D$ where $D$ is a collection of discrete natural numbers. Let, $\hat{m_i}(y)$ denotes the marginal likelihood of $k \in D$. Now select the best possible value of $k$ by
\begin{align*}
\hat{K} = \underset{k \in D} {\mathrm{argmax}} ~\log \hat{m_k}(z).
\end{align*}
Afterwards, in this paper, we named the method of estimation with B-spline basis functions corresponding to the $k=\hat{K}$ to be the empirical Bayes (EB) method. 

Instead of choosing the value of $k$ by the empirical Bayes method, another possibility is to perform model averaging with respect to the posterior distribution of $k$ over $D$. To find the weights of the corresponding models, we calculate the marginal likelihood by the Metropolis-Hastings algorithm for each $k$ following the method described in \cite{Chib2001}. After obtaining the marginal likelihoods for each value of $k$, we obtain the weighted average of the estimated quantile curves where weights are proportional to their marginal likelihoods. Let, $w_k$ and $\hat{m_k}(y)$ denote the weights and the marginal likelihoods for $k \in D$. Let $\hat{\xi}_{1,k}(\tau)$ and $\hat{\xi}_{2,k}(\tau)$ be the posterior means of $\xi_1$ and $\xi_2$ respectively for given $k$. Then we have 
\begin{align*}
w_k &= \frac{\hat{m_k}(y)}{\sum_{k\in D}\hat{m_k}(y)}, \quad k \in D \\
\hat{\xi_1}(\tau) &= \sum_{k \in D}w_k\hat{\xi}_{1,k}(\tau) \quad \text{and} \quad \hat{\xi_2}(\tau) = \sum_{k \in D}w_k\hat{\xi}_{2,k}(\tau).
\end{align*}
Monotonicity of $\hat{\xi}_{1,k}(\tau)$ for each $k \in D$ ensures that $\hat{\xi}_{1}(\tau)$ is also monotonic. Similarly, $\hat{\xi}_{2}(\tau)$ is also monotonic. We note that the estimated quantile functions for each $k$ is totally determined by the estimated coefficients of the corresponding basis functions. Since their relation is also linear, to calculate the overall posterior mean quantile function, it suffices to obtain the posterior mean of the coefficients of the basis functions for each $k$. With the knowledge of weights and posterior mean of basis coefficients for each $k$, we can derive the estimated slope, intercept and quantile functions easily without saving the quantile function values over grid points. For the rest of this paper, we called this to be the hierarchical Bayes (HB) method.

\section{Large-sample Properties}
\label{section_asymptotic}

In this section, we give arguments that indicate the posterior distribution based on a random series of B-splines concentrates near the truth. We assume that the true quantile regression function $Q_0(\tau|x)$ (which necessarily has the linear structure of the form $\beta_{0,\text{true}}(u)+\beta_{\text{true}}(u)x$ is absolutely continuous with respect to the Lebesgue measure and $q_0(\tau|x)$ is the corresponding ``quantile density function''. We assume that $q_0$ satisfies the following regularity condition: 

(A) For any $\epsilon>0$, there exists $\delta>0$ such that for any strictly increasing, continuous, bijection $h:[0,1]\to [0,1]$ such that $\sup_u|h(u)-u|<\delta$, we have $\int \log {q_0(h(u)|x)}{q_0(u|x)}du <\epsilon$ for all $x$. 

We also assume that the true conditional density function $f_0(y|x)$ is positive and continuous (and hence is bounded above and below) and that the distribution of $X$ is $G$. Let $F_0(y|x)$ stand for the true conditional distribution function. 

Clearly the random quantile regression function (i.e., in the model following the B-spline basis expansion prior) $Q(\tau|x)$ is differentiable with quantile density function $q(\tau|x)$. Let the corresponding conditional density function be denoted by $f(y|x)$ and the cumulative distribution function $F(y|x)$. We assume that the prior on $k$ is positive for all values. 

It follows from a well-known theorem of \cite{Schwartz1965} that the posterior distribution of the joint distribution of $(X,Y)$ is consistent with respect to the weak topology if for every $\epsilon>0$, the prior probability of 
\begin{equation}
\int \int f_0(y|x) \log \frac{f_0(y|x)}{f(y|x)} dy \, dG(x)<\epsilon
\label{KLproperty}
\end{equation}
is positive. Below we show that the condition holds at the true $f_0$ for the B-spline random series prior. 

We follow the line of argument given in \cite{Hjort2009} although in our case the presence of the conditioning variable $X$ makes the situation considerably more complicated. Define $h_x(u)=F(Q_0(u|x)|x)$. Note that if $U$ is $U(0,1)$, then $Q_0(U|x)$ has conditional density $f_0(y|x)$. Hence 
\begin{eqnarray}
\lefteqn{ \int \int f_0(y|x) \log \frac{f_0(y|x)}{f(y|x)}dy\, dG(x) }\nonumber\\
&&=\int \int \log \frac{p_0(Q_0(u|x)|x)}{p(Q_0(u|x)|x)}du\, dG(x)\nonumber\\
&&=\int \int \log \frac{q(h_x(u)|x)}{q_0(u|x)}du\, dG(x)\nonumber\\
&&= \int \int \log \frac{q(h_x(u)|x)}{q_0(h_x(u)|x)}du\, dG(x)\nonumber\\
&&\quad + \int \int \log \frac{q_0(h_x(u)|x)}{q_0(u|x)}du\, dG(x),
\label{KL breakup}
\end{eqnarray}
where we have used the relation between quantile density and probability density $q(u|x)=1/f(Q(u|x)|x)$ and $q_0(u|x)=1/f_0(Q_0(u|x)|x)$, and equivalently $f(y|x)=1/q(F(y|x)|x)$ and $f_0(y|x)=1/q_0(F_0(y|x)|x)$. 

Note that the assumption implies that $q_0(u|x)=\beta_{0,\text{true}}'(u)+\beta_{\text{true}}'(u)x$ and $\beta_{0,\text{true}}'(u)$ and $\beta_{\text{true}}'(u)$ are continuous functions, and hence for a sufficiently large number of knots $k$, both functions can be approximated by a linear combinations of B-splines up to any desired degree of accuracy. Fixing $k$ at a sufficiently large value (which has positive prior probability), we find the approximating coefficients, and then consider all vector of coefficients in a small neighborhood around the vector of approximating coefficients. By the choice of continuous and positive prior density on the coefficient vector in the random spline series, it follows that the event has positive prior probability. Thus neighborhoods of the true quantile function get positive prior probabilities. In view of the portmanteau theorem characterizing weak convergence in terms of convergence of quantile functions (see \cite{Vaart1998}, Lemma~21.2), it follows that weak neighborhood of the true distribution get positive prior probabilities, and so do uniform neighborhoods of the true distribution in view of Polya's theorem. This means, in view of the Condition (A) on $q_0$, that the second term in \eqref{KL breakup} can be arbitrarily small with positive probability.  

For the first term in Equation \eqref{KL breakup}, by the same arguments we find that $q$ is uniformly close to $q_0$ with positive probability, and hence the integrand $\log (q(h_x(u)|x)/q_0(h_x(u|x)))$ is small with positive probability since $q_0$ is assumed to be bounded below. 

Thus the prior puts positive mass in neighborhoods defined by Equation \eqref{KLproperty}, and hence by Schwartz's theorem, the posterior probability of all conditional distributions $F(y|x)$ satisfying 
$$\bigg|\int \psi(x,y) dF(y|x)dG(x)-\int  \psi(x,y) dF_0(y|x)dG(x)\bigg|<\epsilon$$ tends to one for any continuous function $\psi$ on $[0,1]\times [0,1]$ and $\epsilon>0$. 

The above conclusion on posterior consistency is about the conditional distribution function in terms of the weak topology, which does not immediately show concentration of posterior for the quantile function in neighborhoods of the true quantile function. To this end, we further assume that the collection of true conditional distribution function $\{F_0(\cdot|x): x\in [0,1]\}$ is equicontinuous, i.e., given $\epsilon>0$, there exists $\delta>0$ such that whenever $|y-y'|<\delta$, we have that $|F_0(y|x)-F_0(y'|x)|<\epsilon$ for all $x$. Then by an easy modification of the proof of Polya's theorem (see Lemma~2.11 of \cite{vanderVaart}), it follows that $\int F_n(y|x)dG(x)\to F_0(y|x)dG(x)$ for all $y$ implies that $\int \sup_{y\in [0,1]}|F_n(y|x)-F_0(y|x)|dG(x)\to 0$. Thus the posterior probability of  $\int \sup_{y\in [0,1]}|F(y|x)-F_0(y|x)|dG(x)<\epsilon$ tends to one for any $\epsilon>0$. Now for any $\tau$, as $F_0(Q_0(\tau|x)|x)=\tau$, 
$$\int |\tau- F_0(Q(\tau|x)|x)|dG(x)=\int |F_0(Q_0(\tau|x)|x)-F_0(Q(\tau|x)|x)|dG(x).$$ 
As $f_0$ is bounded below by a positive number, this implies that the posterior probability of  $\sup_{\tau\in [0,1]}\int |Q(\tau|x)-Q_0(\tau|x)|dG(x)<\epsilon$ tends to one for any $\epsilon>0$.

\section{Simulation Study}
\label{section_simulation}
In Section \ref{section_model}, we note that after some monotonic transformation, every quantile regression function can be represented in the form mentioned in equation (\ref{our_eq}). For simulation purposes, we consider two different true quantile functions structure. For each cases, we compare the true and the estimated values of the slope, intercept and the quantile regression function for $x=0.3, 0.5$ and $0.7$, $\tau=0.25, 0.5$ and $0.75$ for sample size $n=100$ by our proposed methods,  Gaussian SQR (GSQR) proposed in \cite{Tokdar2012}, Bernstein polynomial SQR (BPSQR) proposed in \cite{Reich2011} and {\tt{qreg}} and  {\tt{qreg\_spline}} functions under the {\tt{BSquare}} package in R by \cite{Smith2013} (see \cite{Reich2012},\cite{Reich2013} for details).

For our proposed methods, suppose $0= t_0 < t_1< \ldots < t_k=1$ be the equidistant knots on the interval $[0,1]$ such that $t_i=1/k$ for all $i=0,1,\ldots,k$. Since MCMC does not mix very well for large values of $k$, we considered only first 8 possible values of $k$ for either cases ($D=\{3,4, \ldots, 10\}$ for QSSQR and $D=\{5,6, \ldots, 12\}$ for CSSQR). Separately, for quadratic and cubic B-spline methods, for each of the values of $k$, we run 20000 iterations with 5000 burn-in. For QSSQR, while evaluating likelihood, we can solve (\ref{tau_x_y}) analytically very fast. Although CSSQR can also be solved analytically, we note that the procedure is considerably slower than bisection method. We apply the latter method with precision of $2^{-10} \approx 10^{-3}$, after finding the interval where $y_i$ is located for $i=1,\ldots,n$, by linear search.

To calculate the marginal likelihood of each model, we need to calculate the expression given by Equation (\ref{eq:chib_expression}). To calculate the numerator of that term, we considered the last 5000 iterations and we take $\omega^* = (\{\gamma_j^{(l)}\}_{j=1}^{k+m-1},\{\delta_j^{(l)}\}_{j=1}^{k+m-1})$ for $l=15000$ where $\beta{(l)}$ denotes the $l$-th update of the parameter $\beta$ in MCMC. Thus the numerator term is evaluated by the end of MCMC chain. To calculate the denominator term, we consider 5000 draws of \{$\omega^{(j)}$\} from $q(\omega^*,\omega|y)$. Then we computed the marginal likelihoods of each model. We also computed the estimates corresponding to weighted average quadratic and cubic spline models taking the weights proportional to the marginal likelihoods over the domain of values of $k$. We used parallel computing for different values of $k$ to obtain posterior probabilities.

To implement the GSQR (Gaussian process SQR) method, proposed in \cite{Tokdar2012}, we took $M=101$ equidistant knots over the interval $[0,1]$, the starting and the ending knots being at 0 and 1 respectively. At each iteration step, we save the values of the Gaussian process at those knots and use those values for likelihood evaluations. We ran 20000 iterations with 5000 burn-in. We evaluate the likelihood function by solving (\ref{tau_x_y}) for each data-points using Newton-Raphson method, as mentioned in \cite{Tokdar2012}. We took the precision for convergence criteria to be $10^{-3}$.

While estimating the quantile regression (QRF) functions using the functions {\tt{qreg}} and {\tt{qreg\_spline}} functions under the {\tt{BSquare}} package in R by \cite{BSquare}, for each case we performed 50000 iterations with 10000 burn-in. For BPSQR (Bernstein polynomial SQR) method we performed 10000 iterations with 1000 burn-in. The function (implemented in R by \cite{Reich2011}) for estimating the quantile functions used for this method is available at the link \href{url}{http://www4.stat.ncsu.edu/~reich/code/SpaceQRapprox.R}.

For our proposed method, we calculate the uniform $95 \%$ posterior credible region for the quantile regression function (QRF) $Q(\tau|x),\tau \in [0,1]$ for three distinct co-variate values $x=0.2, 0.5, 0.7$ for $\tau \in [0,1]$ for three different sizes of sample $n=50, 100 \; \text{and} \; 200$. To find the $95 \%$ posterior credible region for the QRF for a given co-variate value using Empirical Bayes method, we calculate the posterior mean of the estimated QRF for all $\tau \in [0,1]$ for each value of $k \in D$ at that given covariate value $X=x$. Since we run $20000$ iterations with burn-in $5000$, only last $H = 15000$ are used for calculating the posterior mean. For a given $k \in D$ and covariate value $X=x$, define $R^{EB}_{(x,k)}=\{a^1_{(x,k)}, \ldots, a^H_{(x,k)}\}$ such that $$a^i_{(x,k)} = \sup_{\tau \in [0,1]}\big|Q^{(i)}(\tau|x,k) - \hat{Q}_y(\tau|x,k)\big|, \, i=1,\ldots,15000$$ where $Q^{(i)}(\tau|x,k)$ denotes the QRF for $X=x$ in the $i$-th iteration after burn-in for B-spline method with $k$ partitions and  $\hat{Q}_y(\tau|x,k)$ denotes the posterior mean of the QRF for $X=x$ for B-spline method with $k$ partitions. After that we calculate the $95^{th}$ percentile of $R^{EB}_{(x,k)}$ for each $k \in D$. Thus we find the size of the $95 \%$ posterior credible region for each value of $k \in D$. To find the coverage, we repeat the simulation study $1000$ times under different random number seeds and count the number of times the distance of the true QRF from the estimated EB estimate for all $\tau \in [0,1]$ is not more than the size of the  $95 \%$ posterior credible region for EB approach of corresponding $k$.
 
In HB method, we again have $20000$ iterations with burn-in $5000$ and only last $H = 15000$ are used for calculating the posterior mean. To find the $95 \%$ posterior credible region for the QRF for a given co-variate value $X=x$ in this case, we calculate $R^{HB}_{(x)} = \{a^1_{(x)}, \ldots, a^H_{(x)}\}$ such that $$a^i_{(x)} = \sup_{\tau \in [0,1]}\big|Q^{(i)}(\tau|x) - \hat{Q}_y(\tau|x)\big|, \, i=1,\ldots,15000$$ where $Q^{(i)}(\tau|x)$ denotes the QRF for $X=x$ in the $i$-th iteration after burn-in for B-spline method and  $\hat{Q}_y(\tau|x)$ denotes the posterior mean of the QRF for $X=x$ for B-spline method. The weights corresponding to each $k \in D$ to find $Q^{(j)}(\tau|x)$ and $\hat{Q}_y(\tau|x)$ are the same and they are derived as mentioned in section  \ref{sec_model_avg}. Then we find the size of the $95 \%$ posterior credible region by calculating the $95^{th}$ percentile of $R^{HB}_{(x)}$. To find the coverage, we repeat the simulation study $1000$ times under different random number seeds and count the number of times the distance of the true QRF from the estimated HB estimate for all $\tau \in [0,1]$ is not more than the size of the  $95 \%$ posterior credible region for HB approach.

It is well-known that posterior credible band of smooth functions have an under coverage property (see \cite{Cox1993}, \cite{Knapik2011},\cite{Szabo2015}, \cite{Yoo2016}). To alleviate the problem, undersmoothing or modifying the credible region is needed. We inflate the obtained credible region by blowing the radius of the region by a slowly increasing factor. We choose the inflation factor $f(n) = 0.8\sqrt{\log n}$ in our examples. This works for all the sample sizes $n=50, 100$ and $200$ under different simulation settings and different true quantile regression functions. We report two simulation studies in this paper described in Sub-section \ref{section_simulation_first} and \ref{section_simulation_second}. We have also provided the posterior coverage with and without inflation of uniform $95 \%$ posterior credible interval for three different sample sizes under two different true quantile regression functions.
\subsection{First Study}
\label{section_simulation_first}
Consider $Q(\tau|x) = x\xi_1(\tau)+(1-x)\xi_2(\tau) $ where
\begin{align*}
\xi_1(\tau) = (1 - A)\tau ^ 2 + A\tau, \; \xi_2(\tau) = (1 - B)\tau ^ 2 + B\tau
\end{align*}
and $A = 0.3, B=0.6$. We note that $\xi_1$ and $\xi_2$ are strictly increasing function from $[0,1]$ to $[0,1]$ satisfying $\xi_1(0) = \xi_2(0) =0$ and $\xi_1(1) = \xi_2(1) =1$. Note then the conditional quantile function is given by $Q(\tau|x) = a(x)\tau^2+b(x)\tau$ where $a(x)=x(1-A)+(1-x)(1-B)$ and $b(x)=xA+(1-x)B$. Observe that since the quantile function is the inverse of the cumulative distribution function, for $U \sim U(0,1)$ the random variable $Q(U|x)$ has conditional quantile function $Q(\tau|x)$. We generate $n$ values $x_1,\cdots, x_n$ of the predictor variable $X$ independently from $U(0,1)$. Then we simulate $Y$ variable from the following equation
\begin{align*}
Y_i=a_iU_i^2+b_iU_i \quad \text{for all} \quad i = 1,\ldots,n
\end{align*}
where $a_i=x_i(1-A)+(1-x_i)(1-B)$, $b_i=x_iA+(1-x_i)B$ and $U_i$'s are i.i.d. $U(0,1)$, $i = 1,\ldots,n$. We simulated $n=100$ observations from the distribution and compared the estimated results with the true ones for both Gaussian and B-spline methods.

\begin{figure}[] 
  \begin{subfigure}[b]{0.5\linewidth}
    \centering
    \includegraphics[width=0.75\linewidth]{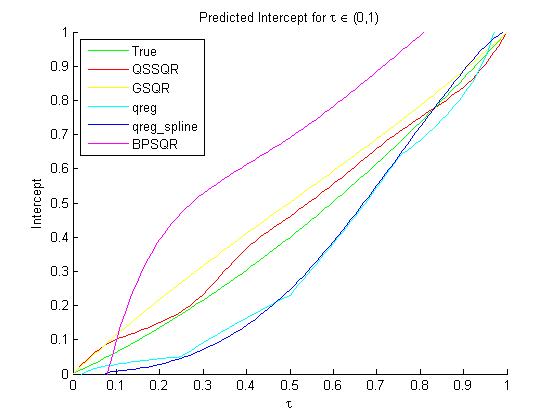} 
    \caption{Intercept} 
    \label{fig:spl_1} 
  \end{subfigure}
  \begin{subfigure}[b]{0.5\linewidth}
    \centering
    \includegraphics[width=0.75\linewidth]{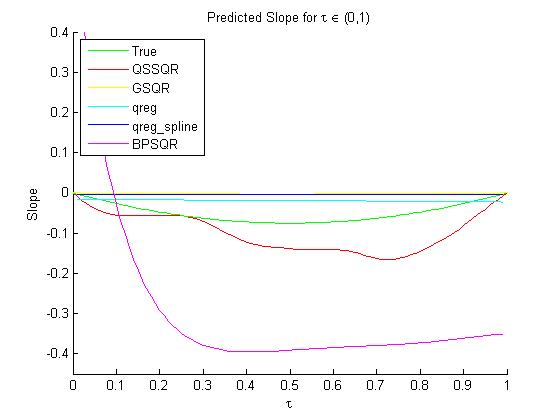} 
    \caption{Slope} 
    \label{fig:spl_2} 
  \end{subfigure} 
  \begin{subfigure}[b]{0.5\linewidth}
    \centering
    \includegraphics[width=0.75\linewidth]{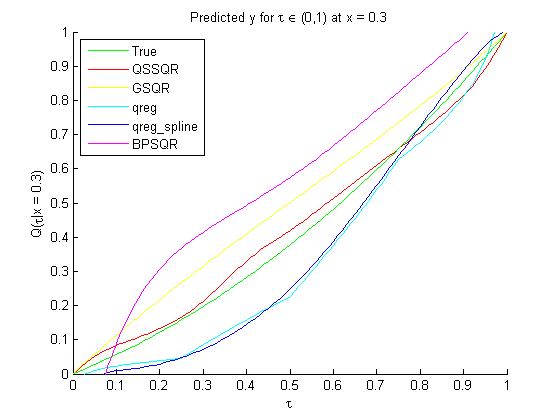} 
    \caption{Estimated QRF at $x = 0.3$} 
    \label{fig:spl_3} 
  \end{subfigure}
  \begin{subfigure}[b]{0.5\linewidth}
    \centering
    \includegraphics[width=0.75\linewidth]{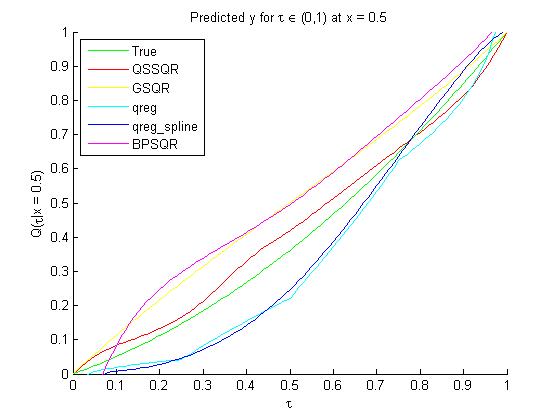} 
    \caption{Estimated QRF at $x = 0.5$} 
    \label{fig:spl_4} 
  \end{subfigure} 
   \begin{subfigure}[b]{0.5\linewidth}
    \centering
    \includegraphics[width=0.75\linewidth]{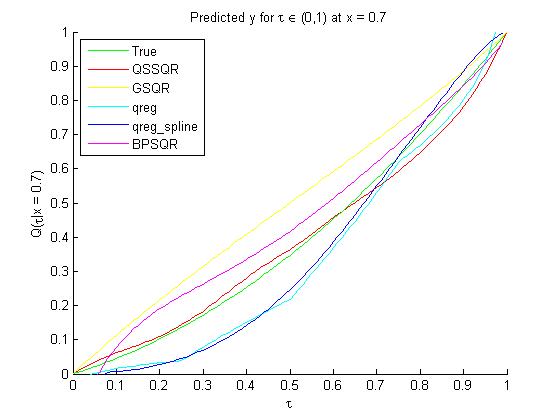} 
    \caption{Estimated QRF at $x = 0.7$} 
    \label{fig:spl_5} 
  \end{subfigure} 
      \begin{subfigure}[b]{0.5\linewidth}
    \centering
    \includegraphics[width=0.75\linewidth]{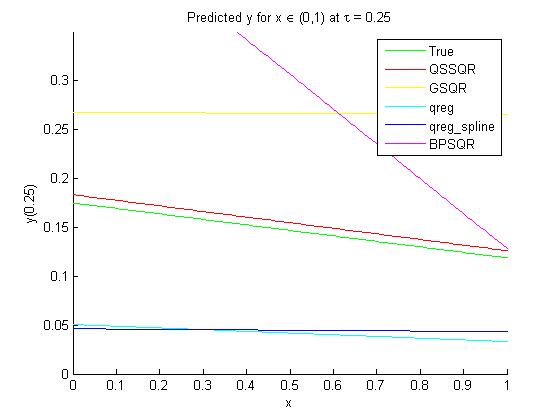} 
    \caption{Estimated QRF at $\tau = 0.25$} 
    \label{fig:spl_6} 
  \end{subfigure}    \begin{subfigure}[b]{0.5\linewidth}
    \centering
    \includegraphics[width=0.75\linewidth]{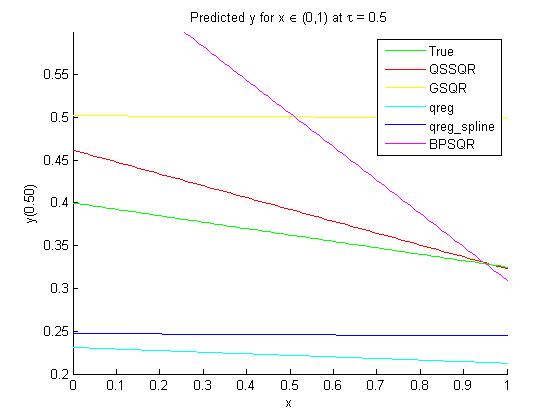} 
    \caption{Estimated QRF at $\tau = 0.5$} 
    \label{fig:spl_7} 
  \end{subfigure}    
  \begin{subfigure}[b]{0.5\linewidth}
    \centering
    \includegraphics[width=0.75\linewidth]{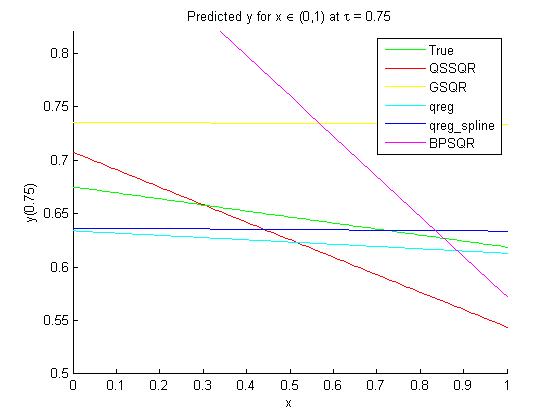} 
    \caption{Estimated QRF at $\tau = 0.75$} 
    \label{fig:spl_8} 
  \end{subfigure} 
  \caption{(First Simulation study) Comparison of true and estimated intercept, slope and estimated quantile regression functions (QRF) at $x=0.3, 0.5, 0.7$, $\tau=0.25, 0.5, 0.75$ and $n=100$ for different methods QSSQR(HB), GSQR, {\tt{qreg}}, {\tt{qreg\_spline}} and BPSQR.}
  \label{figfig} 
\end{figure}

The comparative study of the performances in estimation of our method with other methods under this simulation study has been provided in Figure \ref{figfig}.  We found that there is not much difference between the estimates given by quadratic and cubic B-spline approaches. Hence, for convenience, we only compared the output of our Hierarchical Bayes QSSQR method with other proposed methods in these figures. The root mean integrated squared error (RMISE) is given by the square root of the average of the square of the differences of the estimated and the true values of the curves at those grid points. In Table \ref{table:mse_1}, we compared the RMISE for above mentioned estimated curves for all methods. It can be noted that in our simulation study, the estimated slope, intercept, quantile regression function for $x=0.3, 0.5,0.7$ and $\tau=0.25, 0.5, 0.75$ are curves on the domain $[0,1]$. To calculate the RMISE, we divide the interval $[0,1]$ using partition $(t_0,t_1,\ldots,t_{100})$ such that $0=t_0<t_1<\cdots<t_{100}=1$ such that $(t_i-t_{i-1})=0.01$ for all $i=1,\ldots, 100$. We note that our proposed methods have lower RMISE of estimation curves than that of GSQR. Among other methods, {\tt{qreg}} function from {\tt{BSquare}} worked quite good, though our method yields lower RMISE values for most of the cases. Among our set of proposed methods, we do not see any noticeable improvement by using CSSQR instead of QSSQR. We note that the HB approach has lower RMISE values than the corresponding EB alternative.

\begin{table}[]
\centering
\resizebox{\columnwidth}{!}{%
\bgroup
\def\arraystretch{1}%
\begin{tabular}{|l|c|c|c|c|c|c|c|c|}
\hline
Methods & Intercept & Slope & $x = 0.3$ & $x = 0.5$ & $x = 0.7$ & $\tau = 0.25$ & $\tau = 0.50$ & $\tau = 0.75$ \\ \hline
QSSQR(HB) & 0.039 & 0.059 & 0.029 & 0.037 & 0.030 & 0.008 & 0.035 & 0.038 \\ \hline
QSSQR(EB) & 0.097 & 0.088 & 0.071 & 0.081 & 0.037 & 0.025 & 0.076 & 0.066 \\ \hline
CSSQR(HB) & 0.026 & 0.048 & 0.034 & 0.042 & 0.05 & 0.021 & 0.03 & 0.051 \\ \hline
CSSQR (EB) & 0.033 & 0.069 & 0.034 & 0.041 & 0.051 & 0.051 & 0.051 & 0.014 \\ \hline
GSQR & 0.075 & 0.052 & 0.091 & 0.101 & 0.112 & 0.121 & 0.14 & 0.089 \\ \hline
{\tt{qreg}} & 0.100 & 0.019 & 0.089 & 0.082 & 0.074 & 0.106 & 0.142 & 0.026 \\ \hline
{\tt{qreg\_spline}} & 0.101 & 0.003 & 0.093 & 0.084 & 0.077 & 0.103 & 0.118 & 0.019 \\ \hline
BPSQR & 0.267 & 0.353 & 0.177 & 0.118 & 0.063 & 0.181 & 0.168 & 0.145 \\ \hline
\end{tabular}
\egroup
}
\caption{(First Simulation study) Comparison of RMISE of estimation of slope, intercept and quantile regression function at $x=0.3, 0.5, 0.7$, $\tau = 0.25, 0.50, 0.75$ and $n=100$ under different methods QSSQR(HB); QSSQR(EB); CSSQR(HB); CSSQR(EB); GSQR; {\tt{qreg}}; {\tt{qreg\_spline}} and BPSQR.}
\label{table:mse_1}
\end{table} 

 In Table \ref{table:coverage_1}, we present estimation accuracy and coverage with and without inflation of the proposed methods for three different sample sizes $n=50,100,200$. We note that the posterior coverage of the inflated credible band for the Hierarchical Bayes are typically better than that of corresponding Empirical Bayes. No noticeable improvement is noticed by using CSSQR instead of QSSQR.\\

\begin{table}[]
\centering\resizebox{\columnwidth}{!}{%
\bgroup
\def\arraystretch{0.8}%
\begin{tabular}{|c|c|c|c|c|c|c|c|c|}
\hline
\multirow{2}{*}{Degree} & \multirow{2}{*}{\begin{tabular}[c]{@{}c@{}}Sample\\ size\end{tabular}} & \multirow{2}{*}{Type} & \multicolumn{2}{c|}{$x=0.2$} & \multicolumn{2}{c|}{$x=0.5$} & \multicolumn{2}{c|}{$x=0.7$} \\ \cline{4-9} 
 &  &  & Size & Coverage & Size & Coverage & Size & Coverage \\ \hline
\multirow{6}{*}{QSSQR} & \multirow{2}{*}{$n=50$} & HB & \begin{tabular}[c]{@{}c@{}}\textbf{0.2394}\\ 0.1513\end{tabular} & \begin{tabular}[c]{@{}c@{}}\textbf{99.7}\\ 86.1\end{tabular} & \begin{tabular}[c]{@{}c@{}}\textbf{0.1936}\\ 0.1224\end{tabular} & \begin{tabular}[c]{@{}c@{}}\textbf{99.3}\\ 77.5\end{tabular} & \begin{tabular}[c]{@{}c@{}}\textbf{0.2020}\\ 0.1277\end{tabular} & \begin{tabular}[c]{@{}c@{}}\textbf{98.3}\\ 75.1\end{tabular} \\ \cline{3-9} 
 &  & EB & \begin{tabular}[c]{@{}c@{}}\textbf{0.2327}\\ 0.1471\end{tabular} & \begin{tabular}[c]{@{}c@{}}\textbf{95.9}\\ 70.4\end{tabular} & \begin{tabular}[c]{@{}c@{}}\textbf{0.1935}\\ 0.1223\end{tabular} & \begin{tabular}[c]{@{}c@{}}\textbf{94.2}\\ 63.8\end{tabular} & \begin{tabular}[c]{@{}c@{}}\textbf{0.2210}\\ 0.1397\end{tabular} & \begin{tabular}[c]{@{}c@{}}\textbf{94.4}\\ 67.3\end{tabular} \\ \cline{2-9} 
 & \multirow{2}{*}{$n=100$} & HB & \begin{tabular}[c]{@{}c@{}}\textbf{0.1925}\\ 0.1121\end{tabular} & \begin{tabular}[c]{@{}c@{}}\textbf{96.6}\\ 57.2\end{tabular} & \begin{tabular}[c]{@{}c@{}}\textbf{0.1734}\\ 0.1010\end{tabular} & \begin{tabular}[c]{@{}c@{}}\textbf{98.1}\\ 56.3\end{tabular} & \begin{tabular}[c]{@{}c@{}}\textbf{0.2158}\\ 0.1257\end{tabular} & \begin{tabular}[c]{@{}c@{}}\textbf{99.4}\\ 74.6\end{tabular} \\ \cline{3-9} 
 &  & EB & \begin{tabular}[c]{@{}c@{}}\textbf{0.2391}\\ 0.1393\end{tabular} & \begin{tabular}[c]{@{}c@{}}\textbf{97.2}\\ 67.1\end{tabular} & \begin{tabular}[c]{@{}c@{}}\textbf{0.2079}\\ 0.1211\end{tabular} & \begin{tabular}[c]{@{}c@{}}\textbf{97.8}\\ 66.7\end{tabular} & \begin{tabular}[c]{@{}c@{}}\textbf{0.2270}\\ 0.1322\end{tabular} & \begin{tabular}[c]{@{}c@{}}\textbf{96.6}\\ 67.4\end{tabular} \\ \cline{2-9} 
 & \multirow{2}{*}{$n=200$} & HB & \begin{tabular}[c]{@{}c@{}}\textbf{0.2094}\\ 0.1137\end{tabular} & \begin{tabular}[c]{@{}c@{}}\textbf{98.8}\\ 61.2\end{tabular} & \begin{tabular}[c]{@{}c@{}}\textbf{0.1727}\\ 0.0938\end{tabular} & \begin{tabular}[c]{@{}c@{}}\textbf{98.1}\\ 41.7\end{tabular} & \begin{tabular}[c]{@{}c@{}}\textbf{0.2066}\\ 0.1122\end{tabular} & \begin{tabular}[c]{@{}c@{}}\textbf{99.2}\\ 58.6\end{tabular} \\ \cline{3-9} 
 &  & EB & \begin{tabular}[c]{@{}c@{}}\textbf{0.2296}\\ 0.1247\end{tabular} & \begin{tabular}[c]{@{}c@{}}\textbf{96.9}\\ 65.0\end{tabular} & \begin{tabular}[c]{@{}c@{}}\textbf{0.1803}\\ 0.0979\end{tabular} & \begin{tabular}[c]{@{}c@{}}\textbf{92.7}\\ 47.5\end{tabular} & \begin{tabular}[c]{@{}c@{}}\textbf{0.1994}\\ 0.1083\end{tabular} & \begin{tabular}[c]{@{}c@{}}\textbf{93.7}\\ 51.0\end{tabular} \\ \hline
\multirow{6}{*}{CSSQR} & \multirow{2}{*}{$n=50$} & HB & \begin{tabular}[c]{@{}c@{}}\textbf{0.2475}\\ 0.1564\end{tabular} & \begin{tabular}[c]{@{}c@{}}\textbf{99.8}\\ 86.5\end{tabular} & \begin{tabular}[c]{@{}c@{}}\textbf{0.2109}\\ 0.1333\end{tabular} & \begin{tabular}[c]{@{}c@{}}\textbf{99.7}\\ 78.1\end{tabular} & \begin{tabular}[c]{@{}c@{}}\textbf{0.2225}\\ 0.1406\end{tabular} & \begin{tabular}[c]{@{}c@{}}\textbf{99.8}\\ 78.4\end{tabular} \\ \cline{3-9} 
 &  & EB & \begin{tabular}[c]{@{}c@{}}\textbf{0.2361}\\ 0.1492\end{tabular} & \begin{tabular}[c]{@{}c@{}}\textbf{97.7}\\ 68.7\end{tabular} & \begin{tabular}[c]{@{}c@{}}\textbf{0.1994}\\ 0.1260\end{tabular} & \begin{tabular}[c]{@{}c@{}}\textbf{95.8}\\ 60.7\end{tabular} & \begin{tabular}[c]{@{}c@{}}\textbf{0.1930}\\ 0.1220\end{tabular} & \begin{tabular}[c]{@{}c@{}}\textbf{92.1}\\ 50.7\end{tabular} \\ \cline{2-9} 
 & \multirow{2}{*}{$n=100$} & HB & \begin{tabular}[c]{@{}c@{}}\textbf{0.2316}\\ 0.1349\end{tabular} & \begin{tabular}[c]{@{}c@{}}\textbf{98.9}\\ 68.9\end{tabular} & \begin{tabular}[c]{@{}c@{}}\textbf{0.1775}\\ 0.1034\end{tabular} & \begin{tabular}[c]{@{}c@{}}\textbf{95.9}\\ 41.0\end{tabular} & \begin{tabular}[c]{@{}c@{}}\textbf{0.1952}\\ 0.1137\end{tabular} & \begin{tabular}[c]{@{}c@{}}\textbf{95.7}\\ 49.6\end{tabular} \\ \cline{3-9} 
 &  & EB & \begin{tabular}[c]{@{}c@{}}\textbf{0.2168}\\ 0.1263\end{tabular} & \begin{tabular}[c]{@{}c@{}}\textbf{95.8}\\ 61.5\end{tabular} & \begin{tabular}[c]{@{}c@{}}\textbf{0.1768}\\ 0.1030\end{tabular} & \begin{tabular}[c]{@{}c@{}}\textbf{94.5}\\ 46.3\end{tabular} & \begin{tabular}[c]{@{}c@{}}\textbf{0.1882}\\ 0.1096\end{tabular} & \begin{tabular}[c]{@{}c@{}}\textbf{92.7}\\ 46.0\end{tabular} \\ \cline{2-9} 
 & \multirow{2}{*}{$n=200$} & HB & \begin{tabular}[c]{@{}c@{}}\textbf{0.1978}\\ 0.1074\end{tabular} & \begin{tabular}[c]{@{}c@{}}\textbf{94.8}\\ 34.2\end{tabular} & \begin{tabular}[c]{@{}c@{}}\textbf{0.1749}\\ 0.0950\end{tabular} & \begin{tabular}[c]{@{}c@{}}\textbf{92.7}\\ 20.6\end{tabular} & \begin{tabular}[c]{@{}c@{}}\textbf{0.2088}\\ 0.1134\end{tabular} & \begin{tabular}[c]{@{}c@{}}\textbf{97.8}\\ 40.4\end{tabular} \\ \cline{3-9} 
 &  & EB & \begin{tabular}[c]{@{}c@{}}\textbf{0.1921}\\ 0.1043\end{tabular} & \begin{tabular}[c]{@{}c@{}}\textbf{92.9}\\ 38.7\end{tabular} & \begin{tabular}[c]{@{}c@{}}\textbf{0.1729}\\ 0.0939\end{tabular} & \begin{tabular}[c]{@{}c@{}}\textbf{92.2}\\ 29.2\end{tabular} & \begin{tabular}[c]{@{}c@{}}\textbf{0.1911}\\ 0.1038\end{tabular} & \begin{tabular}[c]{@{}c@{}}\textbf{93.5}\\ 37.6\end{tabular} \\ \hline
\end{tabular}
\egroup
}
\caption{(First Simulation study) Size and posterior coverage of inflated (in bold) and regular uniform $95 \%$ posterior credible interval of estimated quantile regression function for $x=0.2, 0.5, 0.7$ for $\tau \in [0,1]$ for three different sizes of sample $n=50, 100, 200$ for QSSQR(HB), QSSQR(EB), CSSQR(HB) and CSSQR(EB).}
\label{table:coverage_1}
\end{table}

\begin{figure}[] 
  \begin{subfigure}[b]{0.25\linewidth}
    \centering
    \includegraphics[width=0.9\linewidth]{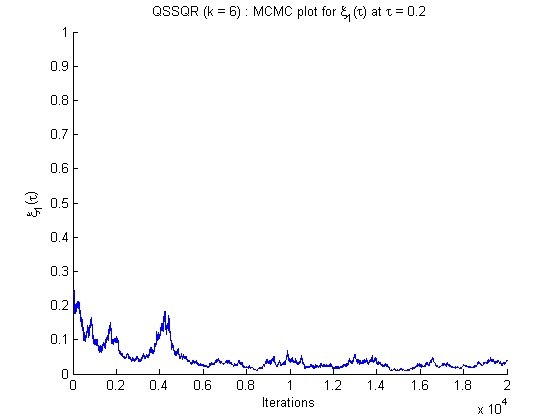} 
    \label{fig:mcmc_1} 
    \caption{QSSQR : $\xi_1(0.2)$}
  \end{subfigure}
  \begin{subfigure}[b]{0.25\linewidth}
    \centering
    \includegraphics[width=0.9\linewidth]{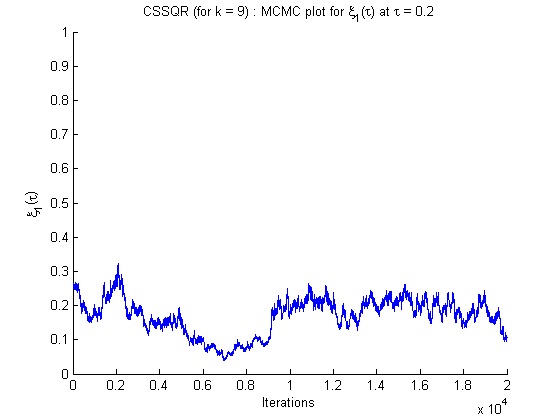}  
    \label{fig:mcmc_2} 
    \caption{CSSQR : $\xi_1(0.2)$}
  \end{subfigure} 
  \begin{subfigure}[b]{0.25\linewidth}
    \centering
    \includegraphics[width=0.9\linewidth]{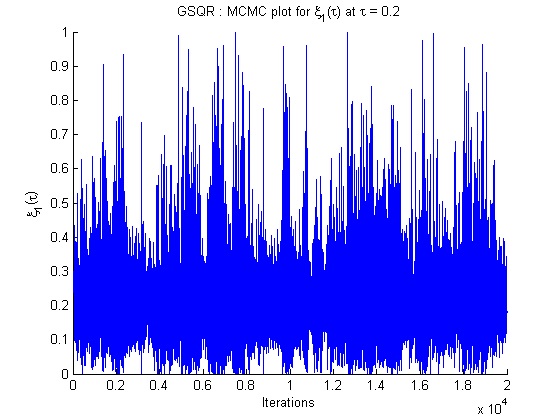} 
    \label{fig:mcmc_3} 
    \caption{GSQR : $\xi_1(0.2)$}
  \end{subfigure}
  \newline
  \begin{subfigure}[b]{0.25\linewidth}
    \centering
    \includegraphics[width=0.9\linewidth]{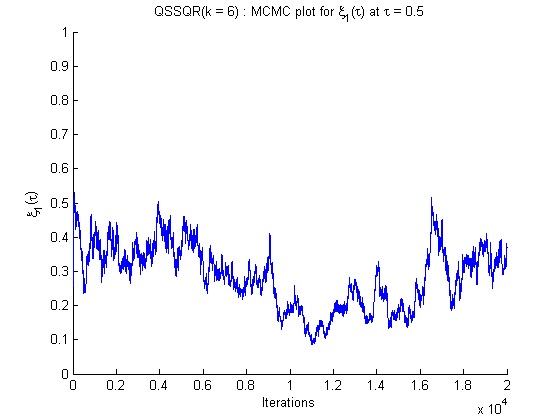} 
    \label{fig:mcmc_4} 
    \caption{QSSQR : $\xi_1(0.5)$}
  \end{subfigure} 
   \begin{subfigure}[b]{0.25\linewidth}
    \centering
    \includegraphics[width=0.9\linewidth]{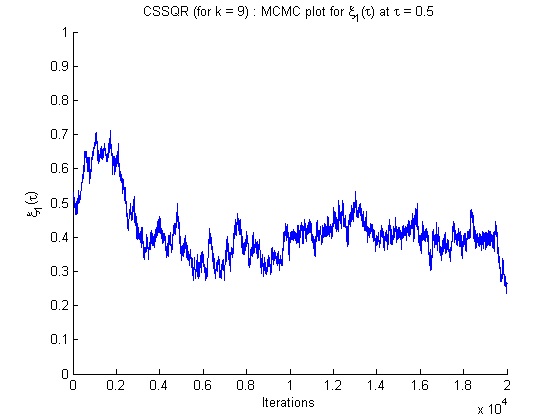} 
    \label{fig:mcmc_5} 
    \caption{CSSQR : $\xi_1(0.5)$}
  \end{subfigure} 
      \begin{subfigure}[b]{0.25\linewidth}
    \centering
    \includegraphics[width=0.9\linewidth]{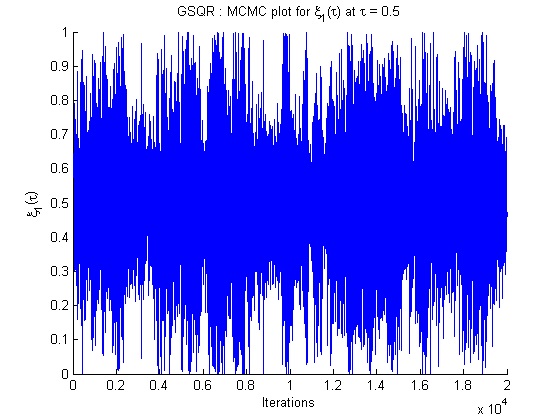}  
    \label{fig:mcmc_6}
    \caption{GSQR : $\xi_1(0.5)$} 
  \end{subfigure}
    \newline
    \begin{subfigure}[b]{0.25\linewidth}
      \centering
      \includegraphics[width=0.9\linewidth]{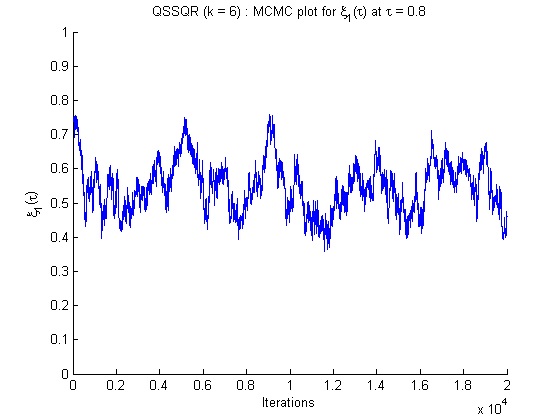} 
      \label{fig:mcmc_7} 
     \caption{QSSQR : $\xi_1(0.8)$}
    \end{subfigure} 
     \begin{subfigure}[b]{0.25\linewidth}
      \centering
      \includegraphics[width=0.9\linewidth]{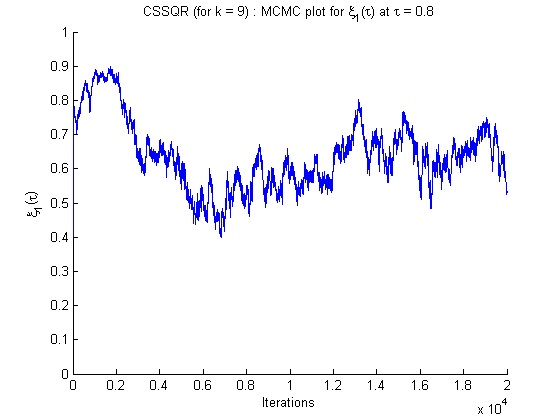} 
      \label{fig:mcmc_8} 
      \caption{CSSQR : $\xi_1(0.8)$}
    \end{subfigure} 
        \begin{subfigure}[b]{0.25\linewidth}
      \centering
      \includegraphics[width=0.9\linewidth]{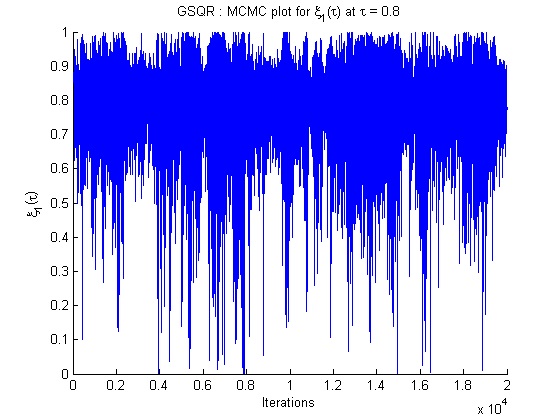}  
      \label{fig:mcmc_9} 
      \caption{GSQR : $\xi_1(0.8)$}
    \end{subfigure} 
          \newline
          \begin{subfigure}[b]{0.25\linewidth}
            \centering
            \includegraphics[width=0.9\linewidth]{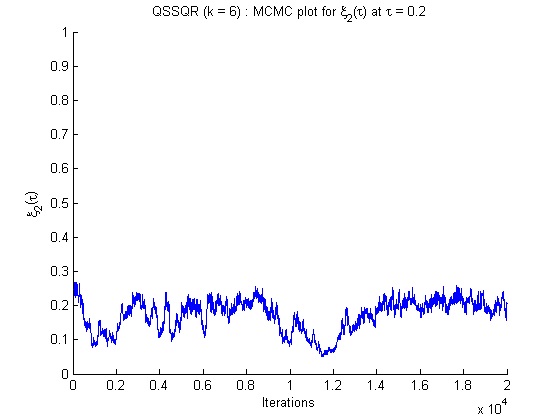} 
            \label{fig:mcmc_10} 
            \caption{QSSQR : $\xi_2(0.2)$}
          \end{subfigure} 
           \begin{subfigure}[b]{0.25\linewidth}
            \centering
            \includegraphics[width=0.9\linewidth]{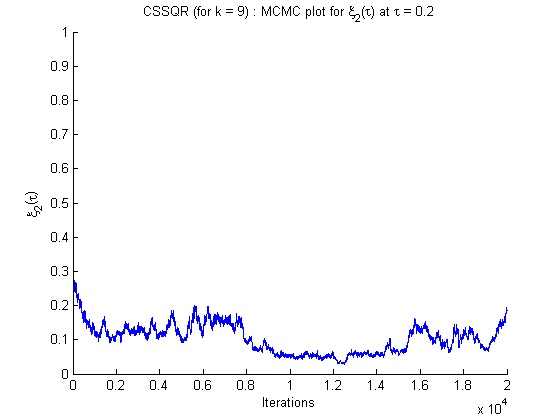} 
            \label{fig:mcmc_11} 
            \caption{CSSQR : $\xi_2(0.2)$}
          \end{subfigure} 
              \begin{subfigure}[b]{0.25\linewidth}
            \centering
            \includegraphics[width=0.9\linewidth]{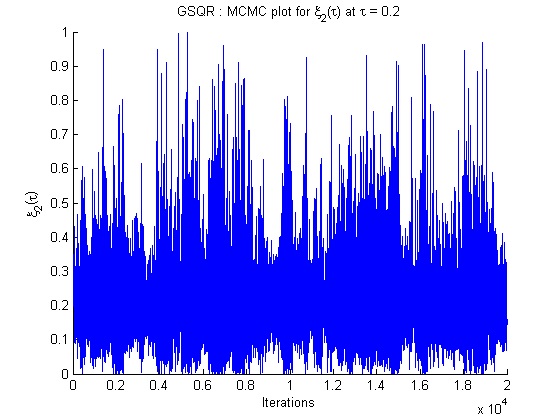}  
            \label{fig:mcmc_12} 
            \caption{GSQR : $\xi_2(0.2)$}
          \end{subfigure} 
              \newline
              \begin{subfigure}[b]{0.25\linewidth}
                \centering
                \includegraphics[width=0.9\linewidth]{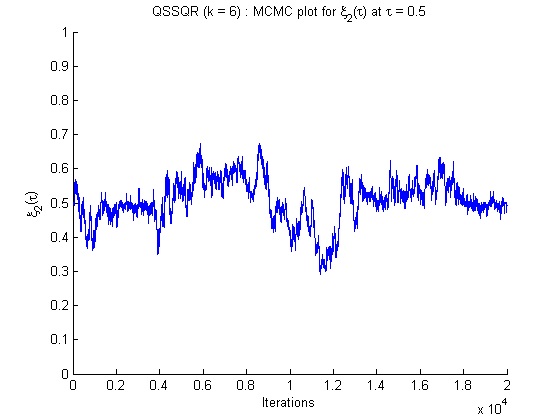} 
                \label{fig:mcmc_13} 
                \caption{QSSQR : $\xi_2(0.5)$}
              \end{subfigure} 
               \begin{subfigure}[b]{0.25\linewidth}
                \centering
                \includegraphics[width=0.9\linewidth]{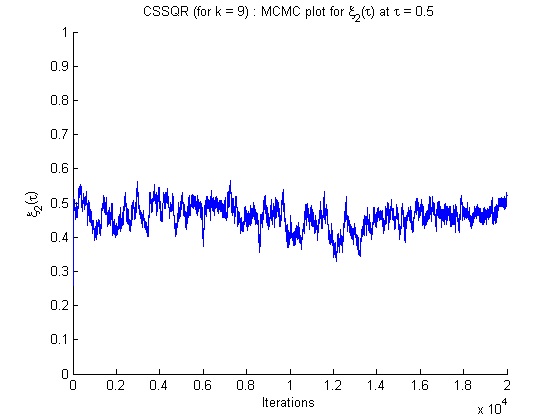} 
                \label{fig:mcmc_14} 
                \caption{CSSQR : $\xi_2(0.5)$}
              \end{subfigure} 
                  \begin{subfigure}[b]{0.25\linewidth}
                \centering
                \includegraphics[width=0.9\linewidth]{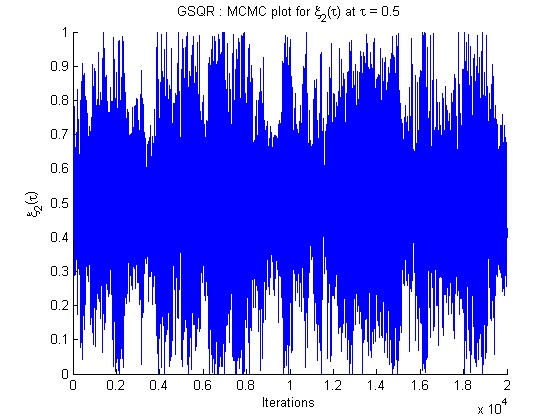}  
                \label{fig:mcmc_15} 
                \caption{GSQR : $\xi_2(0.5)$}
              \end{subfigure} 
                \newline
                \begin{subfigure}[b]{0.25\linewidth}
                  \centering
                  \includegraphics[width=0.9\linewidth]{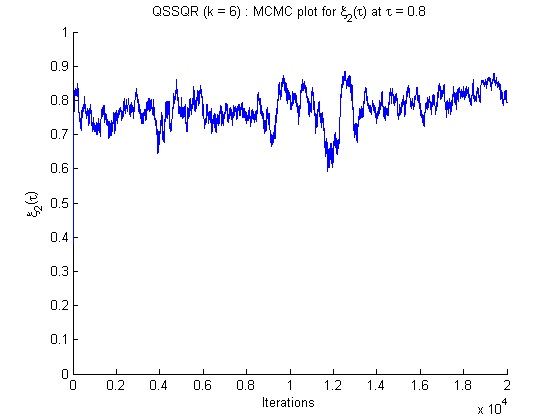} 
                  \label{fig:mcmc_16} 
                 \caption{QSSQR : $\xi_2(0.8)$}
                \end{subfigure} 
                 \begin{subfigure}[b]{0.25\linewidth}
                  \centering
                  \includegraphics[width=0.9\linewidth]{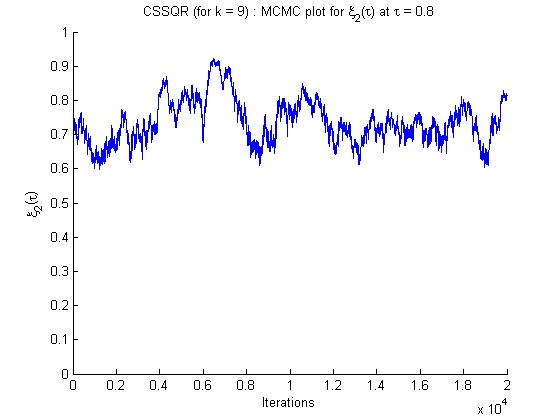} 
                  \label{fig:mcmc_17} 
                  \caption{CSSQR : $\xi_2(0.8)$}
                \end{subfigure} 
                    \begin{subfigure}[b]{0.25\linewidth}
                  \centering
                  \includegraphics[width=0.9\linewidth]{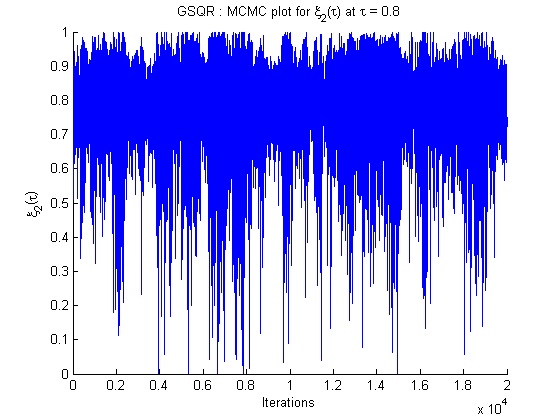}  
                  \label{fig:mcmc_18} 
                  \caption{GSQR : $\xi_2(0.8)$}
                \end{subfigure} 
  \caption{(First Simulation study) MCMC trace plots of $\xi_1(\tau)$ and $\xi_2(\tau)$ at $\tau=0.2,0.5,0.8$ for QSSQR $(k=6)$, CSSQR $(k=9)$ and GSQR.}
  \label{traces} 
\end{figure}

\subsection{Second Study}
\label{section_simulation_second}
To check the performance of the proposed Bayesian method when the quantile function is not a polynomial, we consider,
\begin{align*}
\xi_1(\tau) = \sin(\frac{\pi\tau}{2}), \; \xi_2(\tau) = \frac{\log(1+\tau)}{\log{2}}
\end{align*}
again, we note that $\xi_1$ and $\xi_2$ are strictly increasing function from $[0,1]$ to $[0,1]$ satisfying $\xi_1(0) = \xi_2(0) =0$ and $\xi_1(1) = \xi_2(1) =1$. We generate a sample of size $n=100$ using the quantile function given by 
\begin{align*}
Q(\tau|x) = x\xi_1(\tau)+(1-x)\xi_2(\tau).
\end{align*}
In this case, the assumptions, precision, prior model parameter values, number of iterations and burn-in have been taken to be the same as the previous study for all above-mentioned models. 

\begin{figure}[] 
  \begin{subfigure}[b]{0.5\linewidth}
    \centering
    \includegraphics[width=0.75\linewidth]{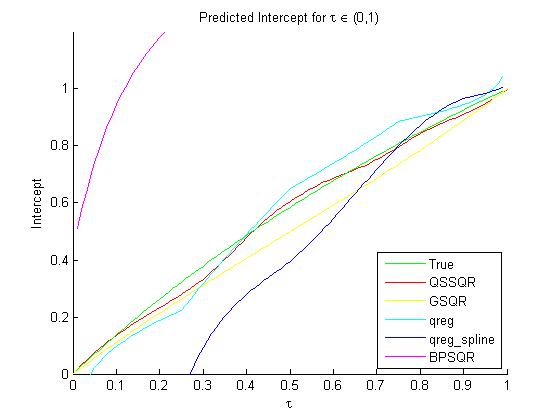} 
    \caption{Intercept} 
    \label{fig:spl_1_2} 
  \end{subfigure}
  \begin{subfigure}[b]{0.5\linewidth}
    \centering
    \includegraphics[width=0.75\linewidth]{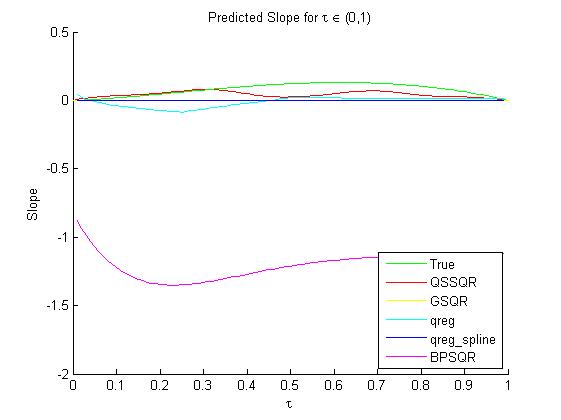} 
    \caption{Slope} 
    \label{fig:spl_2_2} 
  \end{subfigure} 
  \begin{subfigure}[b]{0.5\linewidth}
    \centering
    \includegraphics[width=0.75\linewidth]{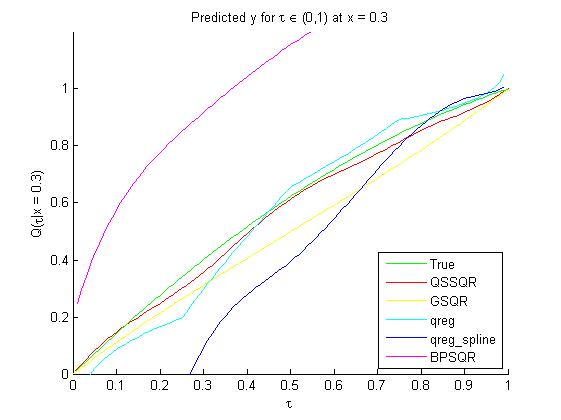} 
    \caption{Estimated QRF at $x = 0.3$} 
    \label{fig:spl_3_2} 
  \end{subfigure}
  \begin{subfigure}[b]{0.5\linewidth}
    \centering
    \includegraphics[width=0.75\linewidth]{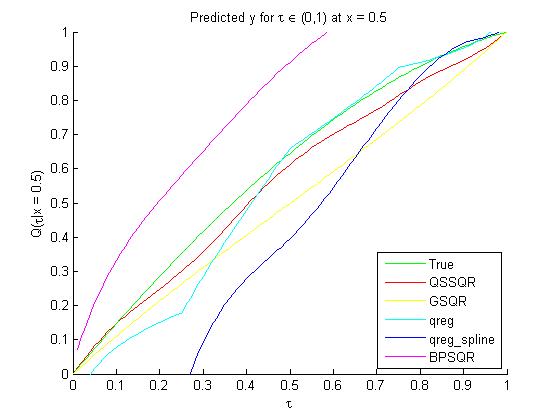} 
    \caption{Estimated QRF at $x = 0.5$} 
    \label{fig:spl_4_2} 
  \end{subfigure} 
   \begin{subfigure}[b]{0.5\linewidth}
   \centering
    \includegraphics[width=0.75\linewidth]{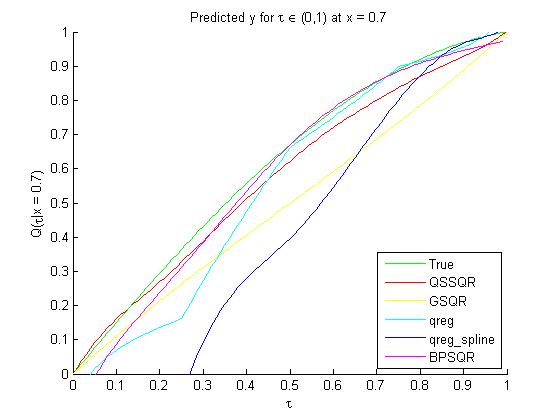} 
    \caption{Estimated QRF at $x = 0.7$} 
    \label{fig:spl_5_2} 
  \end{subfigure}
     \begin{subfigure}[b]{0.5\linewidth}
     \centering
    \includegraphics[width=0.75\linewidth]{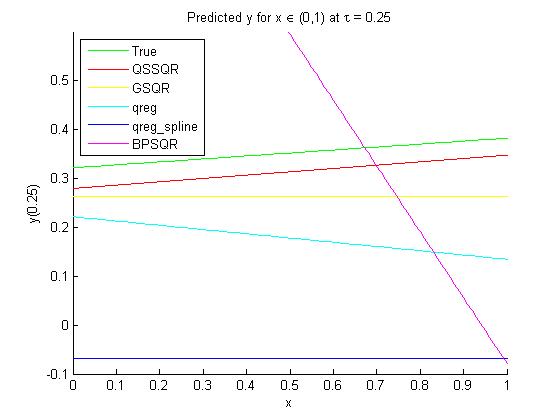} 
    \caption{Estimated QRF at $\tau = 0.25$} 
    \label{fig:spl_6_2} 
  \end{subfigure}
  \begin{subfigure}[b]{0.5\linewidth}
     \centering
    \includegraphics[width=0.75\linewidth]{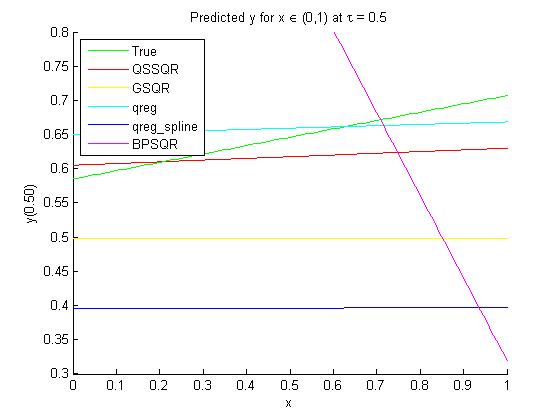} 
    \caption{Estimated QRF at $\tau = 0.50$} 
    \label{fig:spl_7_2} 
  \end{subfigure}
     \begin{subfigure}[b]{0.5\linewidth}
     \centering
    \includegraphics[width=0.75\linewidth]{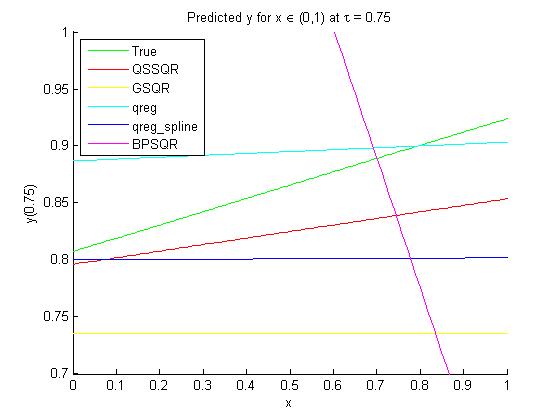} 
    \caption{Estimated QRF at $\tau = 0.75$} 
    \label{fig:spl_8_2} 
  \end{subfigure}
  \caption{(Second Simulation study) Comparison of true and estimated intercept, slope and estimated quantile regression functions (QRF) at $x=0.3, 0.5, 0.7$, $\tau=0.25, 0.5, 0.75$ and $n=100$ for different methods QSSQR(HB), GSQR, {\tt{qreg}}, {\tt{qreg\_spline}} and BPSQR.}
  \label{figfig2} 
\end{figure}

Under this simulation study, the comparative study of the performances in estimation of our method with other methods has been provided in Figure \ref{figfig2}. Like the previous study, in this case also, we found that estimates given by quadratic and cubic B-spline models are similar. So we only compared the output of our QSSQR(HB) with other methods in the above-mentioned figures. In Table \ref{table:mse_2}, we compared the RMISE of the estimated curves for all methods. In Table \ref{table:coverage_2}, estimation performance and posterior coverage with and without inflation using both quadratic and cubic splines are given. In this case also, the proposed methods yield lower RMISE than other methods. Again, no noticeable improvement on using CSSQR over QSSQR has been observed in this scenario also. Like in the previous study, the HB method has lower RMISE values and higher posterior coverage than the corresponding EB method.

In Figure \ref{traces}, the trace plots of $\xi_1(\tau)$ and $\xi_2(\tau)$ have been provided at $\tau=0.2,0.5,0.8$ in the corresponding MCMC using QSSQR, CSSQR and GSQR methods for the first simulation study. For QSSQR and CSSQR, the trace plots have been provided for $k=6$ and $k=9$ respectively. In Table \ref{time}, we compare the computation time of our proposed method with the only existing method of simultaneous quantile regression (if we disregard other existing plug-in type methods of non-crossing quantile regression), i.e., GSQR. For QSSQR, the computation time for both HB and EB are same. The same is also true for CSSQR. We note that, computation time of both QSSQR and CSSQR are lower than that of GSQR, but the computation time of CSSQR is greater than that of QSSQR, as expected. For simulation purposes, all these codes have been written in MATLAB. Simulations have been performed in a cluster with DELL R815 Quad Processor AMD Opteron 16 core 2.3 GHz machines with 512GB RAM, each running 64Bit Fedora Core 20.

\begin{table}[]
\centering
\resizebox{\columnwidth}{!}{%
\bgroup
\def\arraystretch{1}%
\begin{tabular}{|l|c|c|c|c|c|c|c|c|}
\hline
Methods & Intercept & Slope & $x = 0.3$ & $x = 0.5$ & $x = 0.7$ & $\tau = 0.25$ & $\tau = 0.50$ & $\tau = 0.75$ \\ \hline
\begin{tabular}[c]{@{}l@{}}QSSQR(HB)\end{tabular} & 0.020 & 0.053 & 0.024 & 0.040 & 0.040 & 0.039 & 0.040 & 0.044 \\ \hline
\begin{tabular}[c]{@{}l@{}}QSSQR(EB)\end{tabular} & 0.041 & 0.103 & 0.025 & 0.035 & 0.047 & 0.064 & 0.058 & 0.034 \\ \hline
\begin{tabular}[c]{@{}l@{}}CSSQR(HB)\end{tabular} & 0.021 & 0.079 & 0.036 & 0.050 & 0.064 & 0.048 & 0.062 & 0.063 \\ \hline
\begin{tabular}[c]{@{}l@{}}CSSQR(EB)\end{tabular} & 0.034 & 0.079 & 0.040 & 0.051 & 0.063 & 0.058 & 0.034 & 0.060 \\ \hline
GSQR & 0.063 & 0.089 & 0.09 & 0.108 & 0.126 & 0.091 & 0.152 & 0.135 \\ \hline
{\tt{qreg}} & 0.057 & 0.100 & 0.061 & 0.071 & 0.085 & 0.179 & 0.033 & 0.041 \\ \hline
{\tt{qreg\_spline}} & 0.640 & 0.089 & 0.654 & 0.659 & 0.664 & 0.421 & 0.252 & 0.072 \\ \hline
BPSQR & 0.873 & 1.204 & 0.488 & 0.232 & 0.040 & 0.470 & 0.470 & 0.438 \\ \hline
\end{tabular}
\egroup
}
\caption{(Second Simulation study) Comparison of RMISE of estimation of slope, intercept and quantile regression function at $x=0.3, 0.5, 0.7$, $\tau = 0.25, 0.50, 0.75$ and $n=100$ under different methods.  QSSQR(HB); QSSQR(EB); CSSQR(HB); CSSQR(EB); GSQR; {\tt{qreg}}; {\tt{qreg\_spline}} and BPSQR.}
\label{table:mse_2}
\end{table}

\begin{table}[]
\centering
\resizebox{\columnwidth}{!}{%
\bgroup
\def\arraystretch{0.8}%
\begin{tabular}{|c|c|c|c|c|c|c|c|c|}
\hline
\multirow{2}{*}{Degree} & \multirow{2}{*}{\begin{tabular}[c]{@{}c@{}}Sample\\ size\end{tabular}} & \multirow{2}{*}{Type} & \multicolumn{2}{c|}{$x=0.2$} & \multicolumn{2}{c|}{$x=0.5$} & \multicolumn{2}{c|}{$x=0.7$} \\ \cline{4-9} 
 &  &  & Size & Coverage & Size & Coverage & Size & Coverage \\ \hline
\multirow{6}{*}{QSSQR} & \multirow{2}{*}{$n=50$} & HB & \begin{tabular}[c]{@{}c@{}}\textbf{0.2755}\\ 0.1741\end{tabular} & \begin{tabular}[c]{@{}c@{}}\textbf{100}\\ 93.3\end{tabular} & \begin{tabular}[c]{@{}c@{}}\textbf{0.2149}\\ 0.1358\end{tabular} & \begin{tabular}[c]{@{}c@{}}\textbf{99.3}\\ 83.7\end{tabular} & \begin{tabular}[c]{@{}c@{}}\textbf{0.2315}\\ 0.1463\end{tabular} & \begin{tabular}[c]{@{}c@{}}\textbf{98.9}\\ 81.4\end{tabular} \\ \cline{3-9} 
 &  & EB & \begin{tabular}[c]{@{}c@{}}\textbf{0.2514}\\ 0.1589\end{tabular} & \begin{tabular}[c]{@{}c@{}}\textbf{97.6}\\ 76.9\end{tabular} & \begin{tabular}[c]{@{}c@{}}\textbf{0.2157}\\ 0.1363\end{tabular} & \begin{tabular}[c]{@{}c@{}}\textbf{96.5}\\ 69.4\end{tabular} & \begin{tabular}[c]{@{}c@{}}\textbf{0.2483}\\ 0.1569\end{tabular} & \begin{tabular}[c]{@{}c@{}}\textbf{96.2}\\ 70.4\end{tabular} \\ \cline{2-9} 
 & \multirow{2}{*}{$n=100$} & HB & \begin{tabular}[c]{@{}c@{}}\textbf{0.2421}\\ 0.1410\end{tabular} & \begin{tabular}[c]{@{}c@{}}\textbf{99.9}\\ 83.2\end{tabular} & \begin{tabular}[c]{@{}c@{}}\textbf{0.1894}\\ 0.1103\end{tabular} & \begin{tabular}[c]{@{}c@{}}\textbf{99.1}\\ 67.3\end{tabular} & \begin{tabular}[c]{@{}c@{}}\textbf{0.2127}\\ 0.1239\end{tabular} & \begin{tabular}[c]{@{}c@{}}\textbf{98.8}\\ 70.4\end{tabular} \\ \cline{3-9} 
 &  & EB & \begin{tabular}[c]{@{}c@{}}\textbf{0.1997}\\ 0.1163\end{tabular} & \begin{tabular}[c]{@{}c@{}}\textbf{93.4}\\ 55.5\end{tabular} & \begin{tabular}[c]{@{}c@{}}\textbf{0.1815}\\ 0.1057\end{tabular} & \begin{tabular}[c]{@{}c@{}}\textbf{93.8}\\ 53.2\end{tabular} & \begin{tabular}[c]{@{}c@{}}\textbf{0.2349}\\ 0.1368\end{tabular} & \begin{tabular}[c]{@{}c@{}}\textbf{96.8}\\ 65.7\end{tabular} \\ \cline{2-9} 
 & \multirow{2}{*}{$n=200$} & HB & \begin{tabular}[c]{@{}c@{}}\textbf{0.1701}\\ 0.0924\end{tabular} & \begin{tabular}[c]{@{}c@{}}\textbf{95.1}\\ 39.6\end{tabular} & \begin{tabular}[c]{@{}c@{}}\textbf{0.1539}\\ 0.0836\end{tabular} & \begin{tabular}[c]{@{}c@{}}\textbf{94.1}\\ 26.6\end{tabular} & \begin{tabular}[c]{@{}c@{}}\textbf{0.1926}\\ 0.1046\end{tabular} & \begin{tabular}[c]{@{}c@{}}\textbf{98.9}\\ 48.8\end{tabular} \\ \cline{3-9} 
 &  & EB & \begin{tabular}[c]{@{}c@{}}\textbf{0.1921}\\ 0.1043\end{tabular} & \begin{tabular}[c]{@{}c@{}}\textbf{93.8}\\ 51.7\end{tabular} & \begin{tabular}[c]{@{}c@{}}\textbf{0.1856}\\ 0.1008\end{tabular} & \begin{tabular}[c]{@{}c@{}}\textbf{95.1}\\ 50.5\end{tabular} & \begin{tabular}[c]{@{}c@{}}\textbf{0.2175}\\ 0.1181\end{tabular} & \begin{tabular}[c]{@{}c@{}}\textbf{96.4}\\ 58.0\end{tabular} \\ \hline
\multirow{6}{*}{CSSQR} & \multirow{2}{*}{$n=50$} & HB & \begin{tabular}[c]{@{}c@{}}\textbf{0.2563}\\ 0.1620\end{tabular} & \begin{tabular}[c]{@{}c@{}}\textbf{100}\\ 89.5\end{tabular} & \begin{tabular}[c]{@{}c@{}}\textbf{0.1880}\\ 0.1188\end{tabular} & \begin{tabular}[c]{@{}c@{}}\textbf{98.6}\\ 59.4\end{tabular} & \begin{tabular}[c]{@{}c@{}}\textbf{0.1935}\\ 0.1223\end{tabular} & \begin{tabular}[c]{@{}c@{}}\textbf{96.4}\\ 51.4\end{tabular} \\ \cline{3-9} 
 &  & EB & \begin{tabular}[c]{@{}c@{}}\textbf{0.2609}\\ 0.1649\end{tabular} & \begin{tabular}[c]{@{}c@{}}\textbf{98.8}\\ 84.4\end{tabular} & \begin{tabular}[c]{@{}c@{}}\textbf{0.1772}\\ 0.1120\end{tabular} & \begin{tabular}[c]{@{}c@{}}\textbf{89.7}\\ 43.7\end{tabular} & \begin{tabular}[c]{@{}c@{}}\textbf{0.2025}\\ 0.1280\end{tabular} & \begin{tabular}[c]{@{}c@{}}\textbf{89.8}\\ 46.2\end{tabular} \\ \cline{2-9} 
 & \multirow{2}{*}{$n=100$} & HB & \begin{tabular}[c]{@{}c@{}}\textbf{0.2010}\\ 0.1171\end{tabular} & \begin{tabular}[c]{@{}c@{}}\textbf{96.0}\\ 54.2\end{tabular} & \begin{tabular}[c]{@{}c@{}}\textbf{0.1636}\\ 0.0953\end{tabular} & \begin{tabular}[c]{@{}c@{}}\textbf{92.5}\\ 29.9\end{tabular} & \begin{tabular}[c]{@{}c@{}}\textbf{0.2201}\\ 0.1282\end{tabular} & \begin{tabular}[c]{@{}c@{}}\textbf{99.1}\\ 59.8\end{tabular} \\ \cline{3-9} 
 &  & EB & \begin{tabular}[c]{@{}c@{}}\textbf{0.2192}\\ 0.1277\end{tabular} & \begin{tabular}[c]{@{}c@{}}\textbf{96.4}\\ 63.0\end{tabular} & \begin{tabular}[c]{@{}c@{}}\textbf{0.1755}\\ 0.1022\end{tabular} & \begin{tabular}[c]{@{}c@{}}\textbf{92.7}\\ 42.5\end{tabular} & \begin{tabular}[c]{@{}c@{}}\textbf{0.2209}\\ 0.1287\end{tabular} & \begin{tabular}[c]{@{}c@{}}\textbf{97.6}\\ 57.9\end{tabular} \\ \cline{2-9} 
 & \multirow{2}{*}{$n=200$} & HB & \begin{tabular}[c]{@{}c@{}}\textbf{0.1897}\\ 0.1030\end{tabular} & \begin{tabular}[c]{@{}c@{}}\textbf{94.6}\\ 33.0\end{tabular} & \begin{tabular}[c]{@{}c@{}}\textbf{0.1698}\\ 0.0922\end{tabular} & \begin{tabular}[c]{@{}c@{}}\textbf{92.3}\\ 19.6\end{tabular} & \begin{tabular}[c]{@{}c@{}}\textbf{0.2048}\\ 0.1112\end{tabular} & \begin{tabular}[c]{@{}c@{}}\textbf{96.8}\\ 40.4\end{tabular} \\ \cline{3-9} 
 &  & EB & \begin{tabular}[c]{@{}c@{}}\textbf{0.2051}\\ 0.1114\end{tabular} & \begin{tabular}[c]{@{}c@{}}\textbf{95.1}\\ 46.6\end{tabular} & \begin{tabular}[c]{@{}c@{}}\textbf{0.1777}\\ 0.0965\end{tabular} & \begin{tabular}[c]{@{}c@{}}\textbf{92.2}\\ 35.8\end{tabular} & \begin{tabular}[c]{@{}c@{}}\textbf{0.2204}\\ 0.1197\end{tabular} & \begin{tabular}[c]{@{}c@{}}\textbf{97.3}\\ 57.2\end{tabular} \\ \hline
\end{tabular}
\egroup
}
\caption{(Second Simulation study) Size and posterior coverage of inflated (in bold) and regular uniform $95 \%$ posterior credible interval of estimated quantile regression function for $x=0.2, 0.5, 0.7$ for $\tau \in [0,1]$ for three different sizes of sample $n=50, 100, 200$ for QSSQR(HB), QSSQR(EB), CSSQR(HB) and CSSQR(EB).}
\label{table:coverage_2}
\end{table}

\begin{table}[]
\centering
\begin{tabular}{|c|c|c|c|}
\hline
Methods & QSSQR & CSSQR & GSQR \\ \hline
Study 1 & 672 & 776 & 1479 \\ \hline
Study 2 & 666 & 902 & 1566 \\ \hline
\end{tabular}
\caption{Coputation time (in seconds) of QSSQR, CSSQR and GSQR for simulation study 1 and 2 with sample size $n=100$.}
\label{time}
\end{table}

\section{Application to Hurricane Intensity Data}
\cite{Elsner2008} argued that the strongest hurricanes in the North Atlantic basin have gotten stronger over the last couple of decades. We apply our method to the hurricane intensity data in the North Atlantic basin during the period 1981--2006. We use the weighted quadratic spline procedure, i.e., QSSQR(HB). The whole data can be found in the link \url{http://weather.unisys.com/hurricane/atlantic/}.

To use QSSQR(HB), we first mapped the explanatory variable years to the interval $[0,1]$ by change of scale and origin. In this case, we map the year 1981 to 0 and the year 2006 to 1. To map the response variable, the wind speed of the hurricanes at their maximum to the interval $[0,1]$, we assumed that the velocities of the cyclone are coming from a Pareto distribution. The form of the power-Pareto density is given by 
\begin{align}
f(y)=\frac{ak(y/\sigma)^{k-1}}{\sigma(1+(y/\sigma)^k)^{(a+1)}} \quad y>0 \nonumber
\end{align}
Similar to \cite{Tokdar2012} we fix the values of the parameters as $a=0.45$, $\sigma=52$ and $k=4.9$.
The distribution function is given by 
\begin{align}
F(y)=1-\frac{1}{(1+(y/\sigma)^k)^a}
\label{eq:pareto_dist}
\end{align}
Using equation (\ref{eq:pareto_dist}), we transform the hurricane wind speeds to the percentile values. Now, transformed y-values are well in the $[0,1]$ interval and then apply the QSSQR(HB) method. After we evaluate our estimated quantile functions $\hat{\xi}_1(\tau)$ and $\hat{\xi}_2(\tau)$, we find out the slope and intercept at functions of $\tau$. 

After we evaluate them, suppose we want to estimate the wind speed of hurricanes at any particular year of the period 1981--2006, at a given quantile. First we transform the given year to a value well within the interval $[0,1]$ via linear transformation. Then we evaluate the value of transformed wind speeds. After that  we use the inverse transformation of Equation (\ref{eq:pareto_dist}) on estimated transformed speed to find the estimated wind speed at that desired quantile of that year.

\begin{figure}[ht] 
  \begin{subfigure}[b]{0.5\linewidth}
    \centering
    \caption{Comparison of first and last 10 year data}
    \includegraphics[width=0.75\linewidth]{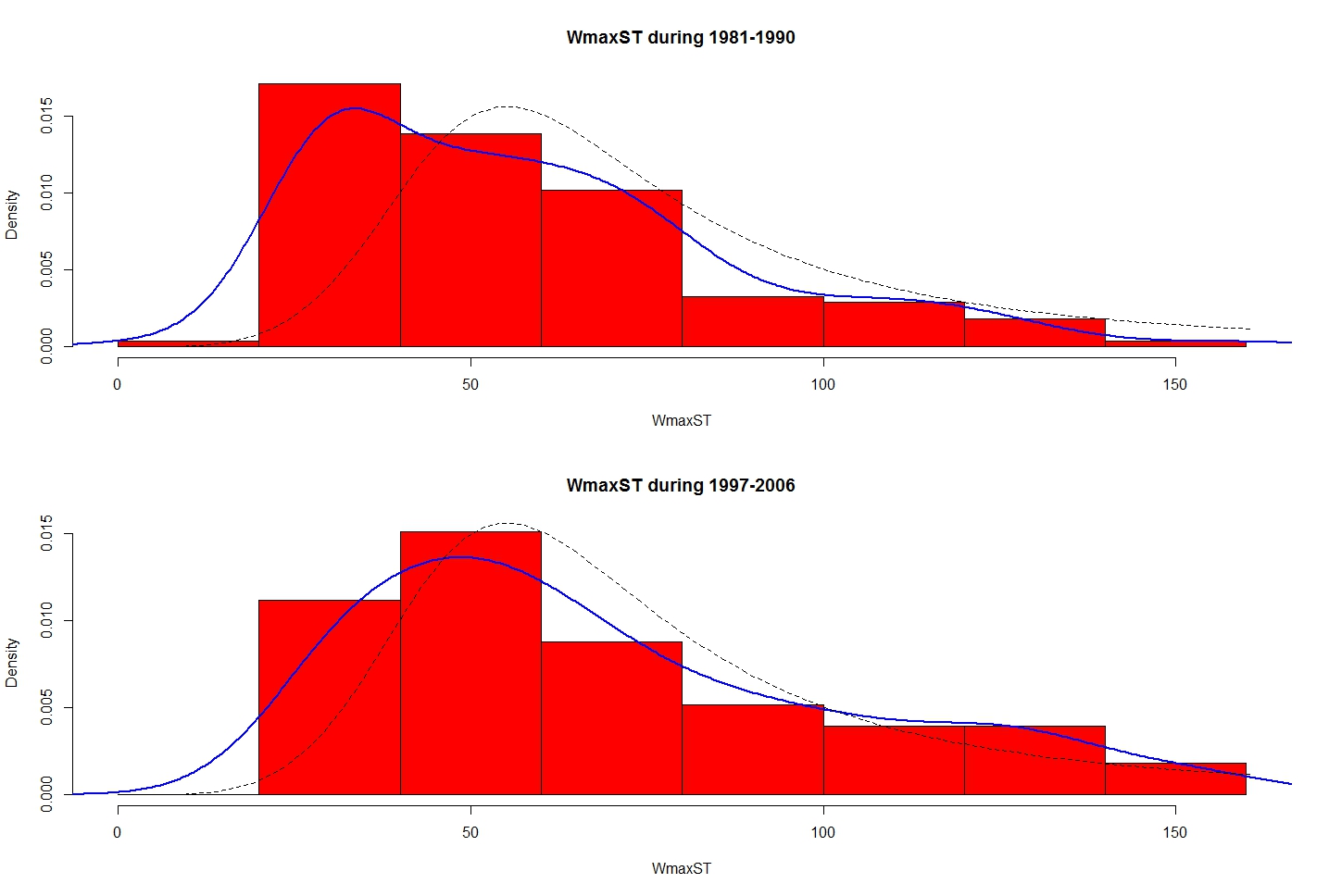} 
    \label{fig:compare_decades} 
    \vspace{4ex}
  \end{subfigure}
  \begin{subfigure}[b]{0.5\linewidth}
    \centering
    \caption{Simultaneous Quantiles}
    \includegraphics[width=0.75\linewidth]{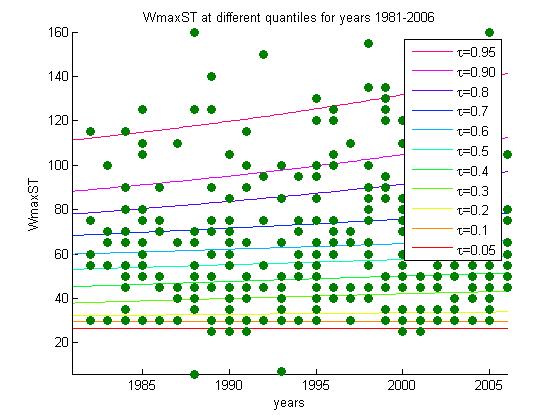} 
    \label{fig:compare_quantiles} 
    \vspace{4ex}
  \end{subfigure} 
  \begin{subfigure}[b]{0.5\linewidth}
    \centering
    \caption{Equidistant year-wise estimation}
    \includegraphics[width=0.75\linewidth]{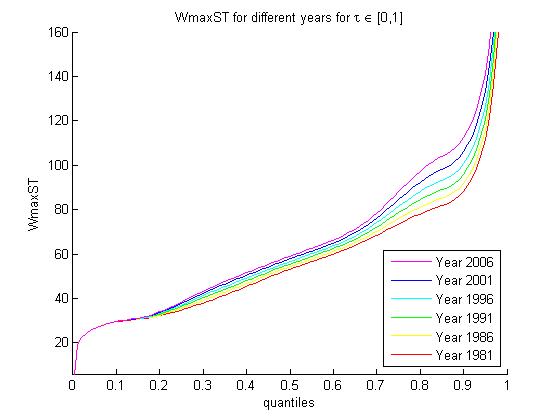} 
    \label{fig:yearly_quantiles} 
  \end{subfigure}
  \begin{subfigure}[b]{0.5\linewidth}
    \centering
    \caption{Probability of negative slope}
    \includegraphics[width=0.75\linewidth]{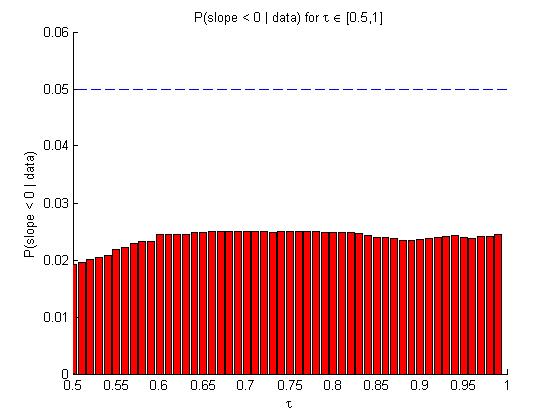} 
    \label{fig:posterior_probs} 
  \end{subfigure} 
  \caption{(a) Comparison of {\tt{WmaxST}} during first and last 10 years : The dotted superimposed line is assumed power-Pareto density used for transforming velocities; the solid superimposed lines are approximate density of {\tt{WmaxST}} for corresponding decades. (b) Simultaneous Quantiles: Estimated quantile curves of velocities over the period 1981--2006 has been shown for the quantiles $\tau \in \{0.05,0.10,0.20,\ldots,0.80,0.90,0.95\}$. (c) {\tt{WmaxST}} as a function of quantiles for the years 1981, 1986, 1991, 1996, 2001 and 2006. (d) Posterior probability of slope being negative for $\tau \in [0.5, 1]$.}
  \label{fig1} 
\end{figure}

In the data, {\tt{WmaxST}} stands for the velocity of the cyclones. In Figure \ref{fig:compare_decades}, we note that the higher velocity cyclones are more frequent in the period 1997--2006 period than 1981--1990 period. We show the estimates for different quantiles over the period 1981--2006 in Figure \ref{fig:compare_quantiles} by using QSSQR (HB). We note that, the the quantile regression curves (QRF) are more steeper for higher quantiles than the lower quantiles.

In order to check whether really the strongest tropical cyclones in the North Atlantic basin have gotten stronger over the last couple of decades (argued by Elsner et al. (2008)), we draw the estimated velocities as a function of quantiles for the years 1981, 1986, 1991, 1996, 2001 and 2006 in Figure \ref{fig:yearly_quantiles}. We note that, for these 6 equidistant years, the estimated {\tt{WmaxST}} corresponding to the lower quantiles are not pretty much different. While, at higher quantiles, the estimated {\tt{WmaxST}} has an increasing trend with time,i.e., more recent years has estimated {\tt{WmaxST}} more than the older years for higher quantiles. In Figure \ref{fig:posterior_probs}, we show the posterior probabilities of the slope to be negative at the higher quantiles, i.e., for $\tau \in [0.5,1]$.

\section{Application to US Population Data}
For last few decades, the population of the states of USA are changing. But the rate of change of population is not same over all zones of USA. We can divide the whole USA mainly in 4 regions namely Northeast, Midwest, South and West. Due to the current trend of globalization, the rate of change of population over the time is different for all these regions of USA. We apply the QSSQR(HB) method of simultaneous quantile regression on the population data of USA over the period 1985--2010. We use the USGS data where we can found population of each county of USA for the years 1985, 1990, 1995, 2000, 2005 and 2010. The whole data can be found in the link \url{http://water.usgs.gov/watuse/data/}.

Before applying our method, we did a monotone transformation so that both predictor and response variables lie in between 0 and 1. We transform the years to the unit interval via linear transformation so that the year 1985 gets mapped to 0 and the year 2010 gets mapped to 1. For the transformation of the population, we considered, the county-wise population of each region follows log-normal density. We fit log-normal density to each of these regions separately. Then for each region, for each county, we use the cumulative distribution function of the population according to the corresponding log-normal distribution. After transforming both explanatory and response variables into the unit interval, we did our analysis. After our analysis we transform the results back to the original scale via the inverse transformation. 

\begin{figure}[ht] 
  \begin{subfigure}[b]{0.5\linewidth}
    \centering
    \includegraphics[width=0.75\linewidth]{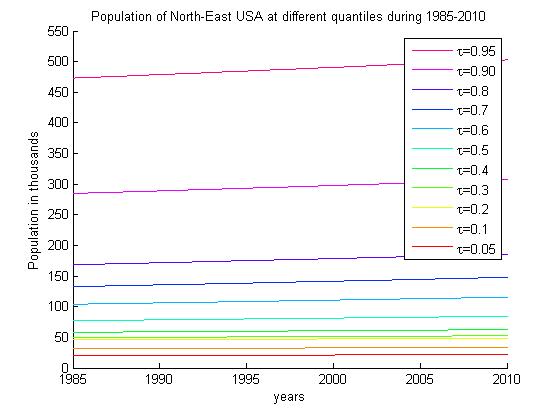} 
    \caption{Northeast} 
    \label{fig1:a} 
    \vspace{4ex}
  \end{subfigure}
  \begin{subfigure}[b]{0.5\linewidth}
    \centering
    \includegraphics[width=0.75\linewidth]{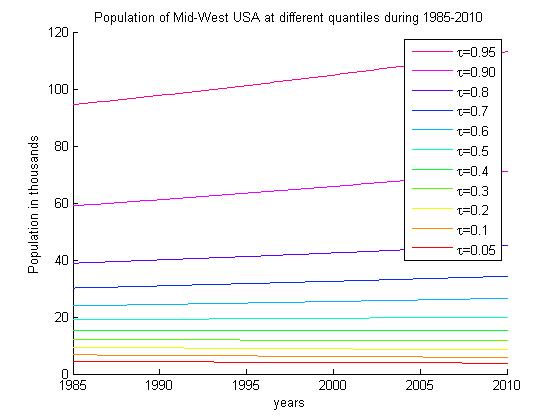} 
    \caption{Midwest} 
    \label{fig1:b} 
    \vspace{4ex}
  \end{subfigure} 
  \begin{subfigure}[b]{0.5\linewidth}
    \centering
    \includegraphics[width=0.75\linewidth]{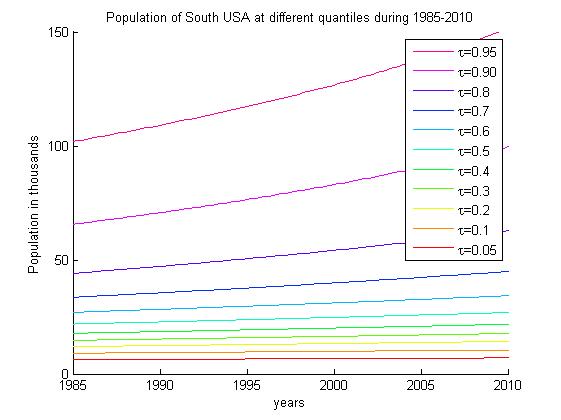} 
    \caption{South} 
    \label{fig1:c} 
  \end{subfigure}
  \begin{subfigure}[b]{0.5\linewidth}
    \centering
    \includegraphics[width=0.75\linewidth]{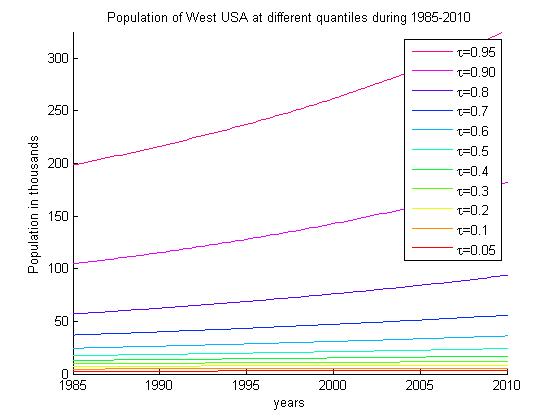} 
    \caption{West} 
    \label{fig1:d} 
  \end{subfigure} 
  \caption{Estimated quantile curves of county wise population of Northeast, Midwest, South and West regions of USA for the quantiles $\tau \in \{0.05,0.10,0.20,\ldots,0.80,0.90,0.95\}$ over the years 1985--2010.}
  \label{fig1} 
\end{figure}

\begin{figure}[ht] 
  \begin{subfigure}[b]{0.5\linewidth}
    \centering
    \includegraphics[width=0.75\linewidth]{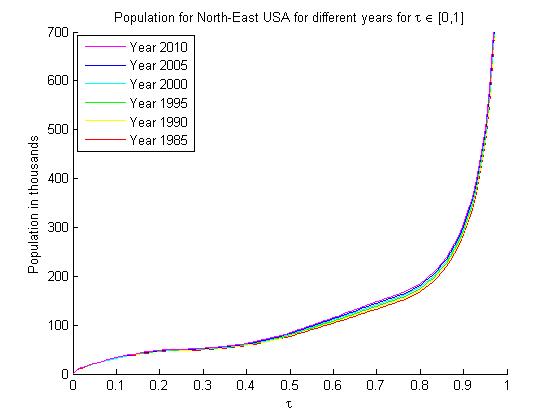} 
    \caption{North-east} 
    \label{fig2:a} 
    \vspace{4ex}
  \end{subfigure}
  \begin{subfigure}[b]{0.5\linewidth}
    \centering
    \includegraphics[width=0.75\linewidth]{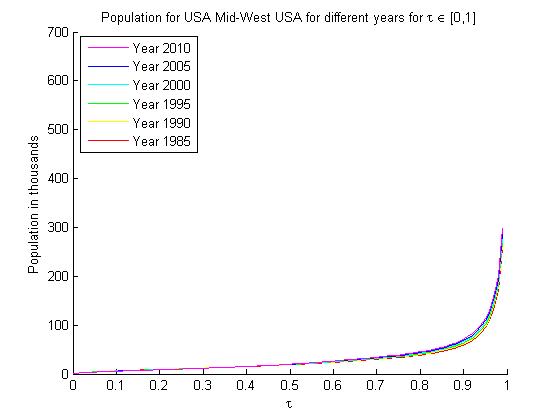} 
    \caption{Mid-west} 
    \label{fig2:b} 
    \vspace{4ex}
  \end{subfigure} 
  \begin{subfigure}[b]{0.5\linewidth}
    \centering
    \includegraphics[width=0.75\linewidth]{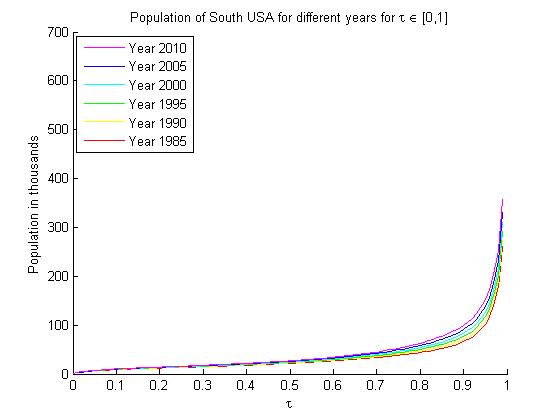} 
    \caption{South} 
    \label{fig2:c} 
  \end{subfigure}
  \begin{subfigure}[b]{0.5\linewidth}
    \centering
    \includegraphics[width=0.75\linewidth]{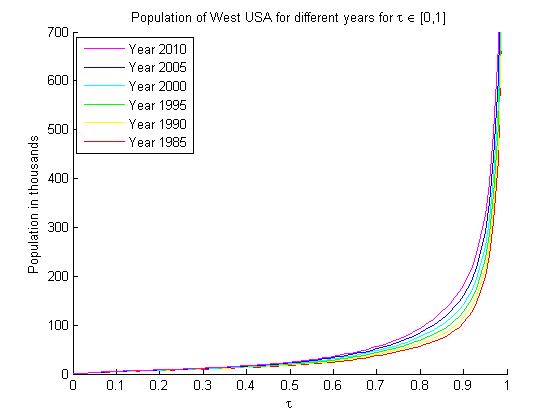} 
    \caption{West} 
    \label{fig2:d} 
  \end{subfigure} 
  \caption{Estimated population for $\tau \in [0,1]$ of Northeast, Midwest, South and West regions of USA for the years 1985, 1990, 1995, 2000, 2005 and 2010}
  \label{fig2} 
\end{figure}

In Figure \ref{fig1}, we show the simultaneous quantiles of county-wise population for all 4 regions of USA over the period 1985--2010. In Figure \ref{fig2} we plotted the estimated quantile regression function curve of population for $\tau \in [0,1]$ for the years 1985, 1990, 1995, 2000, 2005 and 2010 for all regions. 

\begin{figure}[ht] 
  \begin{subfigure}[b]{0.5\linewidth}
    \centering
    \includegraphics[width=0.75\linewidth]{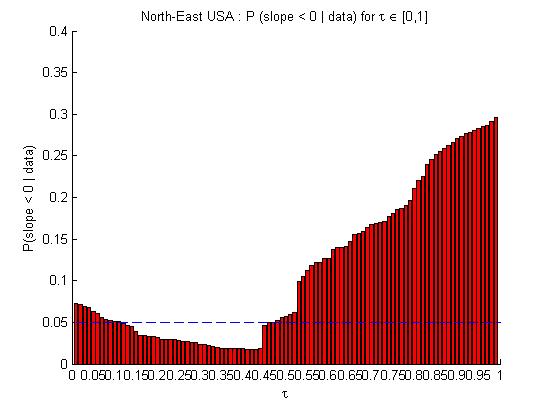} 
    \caption{North-east} 
    \label{fig3:a} 
    \vspace{4ex}
  \end{subfigure}
  \begin{subfigure}[b]{0.5\linewidth}
    \centering
    \includegraphics[width=0.75\linewidth]{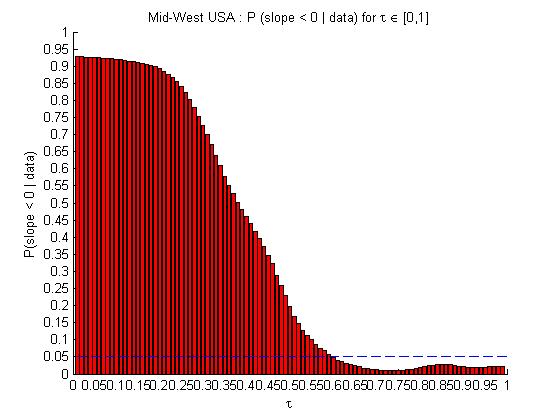} 
    \caption{Mid-west} 
    \label{fig3:b} 
    \vspace{4ex}
  \end{subfigure} 
  \begin{subfigure}[b]{0.5\linewidth}
    \centering
    \includegraphics[width=0.75\linewidth]{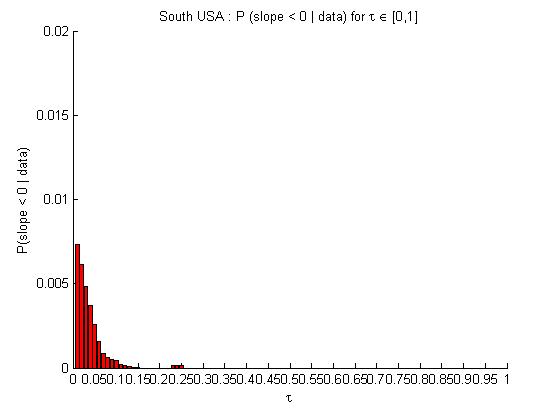} 
    \caption{South} 
    \label{fig3:c} 
  \end{subfigure}
  \begin{subfigure}[b]{0.5\linewidth}
    \centering
    \includegraphics[width=0.75\linewidth]{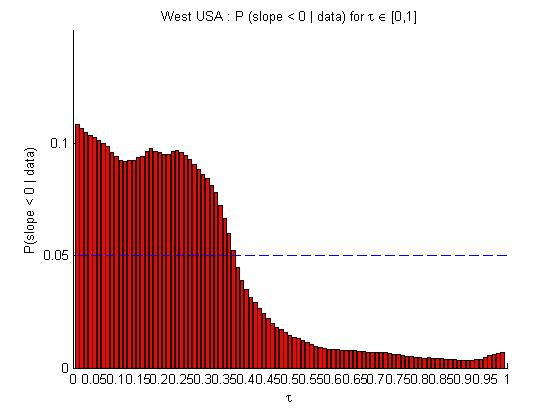} 
    \caption{West} 
    \label{fig3:d} 
  \end{subfigure} 
  \caption{Estimated probability of negative slope for $\tau \in [0,1]$ of Northeast, Midwest, South and West regions of USA for the years 1985, 1990, 1995, 2000, 2005 and 2010}
  \label{fig3} 
\end{figure}

In Figure \ref{fig3} we plot the posterior probabilities of the slope of the simultaneous quantile regression estimates to be negative for $\tau \in [0,1]$ for all 4 regions. We note that for the South region, the posterior probability of the slope being negative is below 0.05 for all quantiles. Therefore in this region, population of all types of counties have increased over the time. In the West, for $\tau$ greater than 0.35, the posterior probability of slope being negative is less than 0.05. That implies population of the more populated counties have increased at higher rate than less populated counties. In the Midwest, we note there is a big difference between the posterior probability of the slope being negative among the sparsely and highly populated counties. Here, the population of sparsely populated counties have decreased over years, on the other hand, the population of highly populated counties have increased. The possible reason is that as fewer and fewer people are now relying on agriculture for their livelihood, new jobs in the Midwest nowadays are available almost exclusively in urban areas. In the Northeast, the probability of the slope being negative is low for sparsely populated counties and relatively high for the highly populated counties. Since there are many highly populated cities in the Northeast, and most of these cities are over-populated and expensive, people tend to move to the outlying areas which still have all facilities and good connection to urban centers. Most people immigrating to USA move in urban centers where jobs are more easily available. 

Overall, we note that except in the Northeast, population of all other regions are significantly increasing for the high-populated counties.  Overall, immigration is a significant force for the population growth of USA. Due to globalization, people from different countries move in USA. Commonly they move in high population areas in the urban areas because offices, business organizations,  universities and other places of interests are mainly located in those regions. This explains the reason why the population of the counties with high population are generally increasing faster over time compared with sparsely populated counties.

\section{Conclusion}
In this paper we have proposed a Bayesian semi-parametric method for fitting simultaneous linear quantile regression using quadratic and cubic B-spline basis function. In order to estimate the coefficients of the B-spline basis functions, we use Metropolis-Hastings algorithm. Finally, to select the optimum number of B-spline basis functions, we calculate the model marginal likelihoods and choose the model with the highest marginal log-likelihood, which we refer as EB. An alternative approach is to use the hierarchical Bayes approach, denoted by HB,  which is a combination of a reasonable range of numbers of B-spline basis function with weights proportional to their model posterior probabilities. From the simulation study on RMISE and posterior coverage of the credible bands, we note somewhat higher accuracy with HB approach over the EB. Since, computation time for both of these methods are the same, HB is preferred over the EB. Even though the use of cubic B-spline basis function is more time consuming than that of the quadratic B-spline basis, we do not find any noticeable improvement in accuracy by using the former over the latter. However the cubic B-spline approach produces estimates which are second order continuously differentiable, instead of only continuously differentiable estimates obtained by the quadratic B-spline approach. 

Unlike most of the previous works on non-crossing quantile regression, the proposed method can estimate the slope and the intercept of the quantile regression equation as continuous function of $\tau$. Most existing methods for non-crossing quantile regression depend on the chosen grid of $\tau$ values where the quantile regression coefficients are estimated. Due to the use of characterization of required monotonicity of the simultaneous linear quantile regression, it can be avoided in our approach. Besides, estimating the quantile regression equation simultaneously gives further insight on the dependence structure of the predictor and the response variable instead of looking at a fixed and finite number of quantiles. 

For the B-spline basis expansion approach, in the likelihood evaluation step, due the piece-wise polynomial structure of B-spline basis function, we can solve the Equation (\ref{tau_x_y}) analytically unlike using the Gaussian process prior where only a numerical solution can be obtained after implementing numerical integration based on a chosen grid. In our proposed method, dependency structure of the slope, intercept and the quantile function of the regression equation with the spline coefficients being linear, it is enough to track the posterior mean of the spline coefficients for the estimation purpose. Evaluation and storage of the estimated quantile functions at grid points over the domain of estimation is unnecessary for our method since, those can be directly derived from the posterior mean of the estimated spline coefficients. So we can unfold each and every details of our estimation by only using posterior mean of a few number of parameters (and model weight vector for HB). This is sharply in contrast of using a Gaussian process prior, for which each realization of the quantile function from its posterior distribution needs to be stored on a grid. The grid also needs to be sufficiently fine to maintain accuracy and smoothness. Though taking more dense grid would improve the quality of estimation, on the other hand, it would increase the computation time and will lead to the problems of singularity of co-variance matrix of the underlying Gaussian process. Again for estimation, we need to evaluate the estimated quantile function at any point by kriging or interpolation. B-spline method does not need any subsequent kriging or interpolation step as the entire estimation is obtained from the estimated coefficients. In our simulation study, we noted that the proposed method based on B-spline basis expansion has slightly more accuracy than the Gaussian process based method for simultaneous quantile regression and several other non-crossing quantile regression methods. 

Application of our method on the north Atlantic hurricane intensity data reveals that the intensity of the strongest hurricanes in the north Atlantic basin have gotten more stronger whereas we did not notice any significant increase in the wind speeds of the hurricanes near median or lower quantiles over the time period 1981--2006. We also applied this method to analyze the rate of change of county level population in the 4 regions of USA, namely, the Northeast, Midwest, South and West, based on the 5-yearly data over the period 1985--2010. It gives us a broader look into how the county population of different regions of USA are changing over time. We note that except sparsely populated counties of the Midwest, the population has generally increased over time in all regions.

\newpage
\section{APPENDIX}
\subsection{Derivation of MCMC Transition Step}
We take the uniform prior on $\{\gamma_j\}_{j=1}^k$ and $\{\delta_j\}_{j=1}^k$ (i.e., Dirichlet(1,\ldots,1)). We generate sequences $\{U_j\}_{j=1}^k$ and $\{W_j\}_{j=1}^k$ of independent random variables from $U(1/r,r)$ for some $r>1$. Since $\{\gamma_j\}_{j=1}^k$ and $\{\delta_j\}_{j=1}^k$ are independent and $\{U_j\}_{j=1}^k$ and $\{W_j\}_{j=1}^k$ are also independent, it is enough to show the results for the first variable transition steps and following conditionals. For the second set of variables, the results will follow similarly. Below $p$ will stand for a generic (joint) density. 

Define $V_j=\gamma_j U_j,\; j=1,\ldots,k$. So we have,
\begin{eqnarray}
p(V_1,\ldots,V_k|\gamma_1,\ldots,\gamma_k)=\prod\limits_{j=1}^k \bigg\{\frac{r}{(r^2-1)\gamma_j}\bigg\} \ I\bigg[\frac{\gamma_j}{r}\leq V_j \leq r\gamma_j\bigg] \nonumber
\end{eqnarray}
Now, define $V=\sum\limits_{j=1}^k V_j.$ After transforming the variables we get, 
\begin{eqnarray}
\lefteqn{p(V_1,\ldots,V_{k-1},V|\gamma_1,\ldots,\gamma_k)}\nonumber\\
&&=\bigg(\frac{r}{r^2-1}\bigg)^k \bigg(\prod\limits_{j=1}^k \gamma_j\bigg)^{-1}\prod\limits_{j=1}^{k-1} I\bigg[\frac{\gamma_j}{r}\leq V_j \leq r\gamma_j\bigg]\nonumber\\ &&\quad \times I\bigg[\frac{\gamma_k}{r} \leq V-\sum\limits_{j=1}^{k-1}V_j \leq r\gamma_k\bigg].\nonumber
\end{eqnarray}
We define $\gamma_j^*=V_j/V, \; j=1,\ldots,k-1$ and set $\gamma_J=(1-\sum\limits_{j=1}^{k-1} \gamma_j)$. Then after transformation of variables we get 
\begin{eqnarray}
\lefteqn{p(\gamma_1^*,\ldots,\gamma_{k-1}^*,V|\gamma_1,\ldots,\gamma_k)}&&\nonumber\\ &&=\bigg(\frac{r}{r^2-1}\bigg)^k V^{k-1} \bigg(\prod\limits_{j=i}^k\gamma_j\bigg)^{-1}\prod\limits_{j=1}^{k-1} I\bigg[\frac{\gamma_j}{rV} \leq \gamma_j^*\leq\frac{r\gamma_j}{V}\bigg] \nonumber \\ 
&&\quad\times I\bigg[\frac{\gamma_k}{r(1-\sum\limits_{j=1}^{k-1} \gamma_j^*)} \leq V \leq \frac{r\gamma_k}{(1-\sum\limits_{j=1}^{k-1} \gamma_j^*)}\bigg]\nonumber\\ &&= \bigg(\frac{r}{r^2-1}\bigg)^J V^{k-1} \bigg(\prod\limits_{j=i}^k\gamma_j\bigg)^{-1}\prod\limits_{j=1}^k I\bigg[\frac{\gamma_j}{rV} \leq \gamma_j^*\leq\frac{r\gamma_j}{V}\bigg]\nonumber\\
&&=\bigg(\frac{r}{r^2-1}\bigg)^k V^{k-1} \bigg(\prod\limits_{j=i}^k\gamma_j\bigg)^{-1}\prod\limits_{j=1}^k I\bigg[\frac{\theta_j}{r\gamma_j^*} \leq V\leq\frac{r\gamma_j}{\gamma_j^*}\bigg]\nonumber\\
&&=\bigg(\frac{r}{r^2-1}\bigg)^k V^{k-1} \bigg(\prod\limits_{j=i}^k\gamma_j\bigg)^{-1}I\bigg[\max_{0 \leq j \leq k}\frac{\gamma_j}{r\gamma_j^*} \leq V\leq \min_{0 \leq j\leq k} \frac{r\gamma_j}{\gamma_j^*}\bigg]. \nonumber
\end{eqnarray}
Hence integrating over the range of $V$, we get 
\begin{align*}
p(\gamma_1^*,\ldots,\gamma_{k-1}^*|\gamma_1,\ldots,\gamma_k)=\int_{\max_{0 \leq j \leq k} \gamma_j/r\gamma_j^*}^{\min_{0 \leq j \leq k} r\gamma_j/\gamma_j^*} \bigg(\frac{r}{r^2-1}\bigg)^k \bigg(\prod\limits_{j=1}^k \gamma_j\bigg)^{-1}V^{k-1}\mathrm{d}V
\end{align*}
Thus we get the conditional density to be
\begin{align*}
p(\gamma^*|\gamma)=\bigg(\frac{r}{r^2-1}\bigg)^k\bigg(\prod\limits_{j=1}^k \gamma_j\bigg)^{-1}\bigg[\Big\{\min_{0 \leq j \leq k}\ \frac{r\gamma_j}{\gamma_j^*}\Big\}^k-\Big\{\max_{0 \leq j \leq k} \frac{\gamma_j}{r\gamma_j^*}\Big\}^k\bigg]\bigg/k.
\end{align*}
Similarly, we obtain,
\begin{align*}
p(\delta^*|\delta)=\bigg(\frac{r}{r^2-1}\bigg)^k\bigg(\prod\limits_{j=1}^k \delta_j\bigg)^{-1}\bigg[\Big\{\min_{0 \leq j \leq k}\ \frac{r\delta_j}{\delta_j^*}\Big\}^k-\Big\{\max_{0 \leq j \leq k} \frac{\delta_j}{r\delta_j^*}\Big\}^k\bigg]\bigg/k.
\end{align*}

\end{document}